\documentclass[a4paper,11pt]{article}
\pdfoutput=1
\usepackage{jcappub} 

\usepackage[T1]{fontenc} 

\usepackage{epsfig}
\usepackage[english]{babel}

\usepackage{amsfonts}
\usepackage{amssymb}
\usepackage{amsmath}
\usepackage{slashbox}
\usepackage{multirow}
\usepackage{comment}

\def\a{\alpha}
\def\b{\beta}
\def\D{\Delta}

\def\l{\lambda}

\def\O{\Phi}

\def\W{\Omega}

\def\z{\zeta}

\def\px{\approx}

\def\={\nonumber &=}
\def\nn{\nonumber}
\def\&{{}&}

\def\({\left(}
\def\){\right)}
\def\[{\left[}
\def\]{\right]}
\def\<{\left\langle}
\def\>{\right\rangle}

\def\un{{\bf \hat{n}}}

\def\curl{\mathcal}

\def\eq{\begin{eqnarray}}
\def\qe{\end{eqnarray}}

\def\and{\quad \mbox{and} \quad}

\def\fnl{f_\textrm{NL}}

\def\Fnl{ F_\textrm {NL}}
\def\barFnl{ \bar F_\textrm {NL}}
\def\Fnllocal{F_\textrm {NL}^\textrm{loc}}
\def\fnllocal{f_\textrm {NL}^\textrm{loc}}

\def\fnlth{f_\textrm {NL}^\textrm{th}}
\def\bfnl{\kern2pt\overline{\kern-2ptf}_\textrm{NL}}

\def\lmax{l_\textrm{max}}
\def\lall{l_1,l_2,l_3}
\def\lalltwo{l_4,l_5,l_6}
\def\lsum{l_1+l_2+l_3}

\def\Blll{B_{l_1l_2l_3}}
\def\blll{b_{l_1l_2l_3}}

\def\hlll{h_{l_1l_2l_3}}
\def\hllltwo{h_{l_4l_5l_6}}

\def\kall{k_1,k_2,k_3}

\def\alm{a_{lm}}

\def\almfour{a_{l_4m_4}}
\def\almfive{a_{l_5m_5}}
\def\almsix{a_{l_6m_6}}

\def\Vtetra{{{\cal V}_{\cal T}}}

\def\Qn{\curl{Q}_n}

\def\Rp{\curl{R}_p}

\def\barQ{\kern2pt\overline{\kern-2pt\curl{Q}}}
\def\barQn{\barQ_n}

\def\barQp{\barQ_p}

\def\barR{\kern2pt\overline{\kern-2pt\curl{R}}}
\def\barRn{\barR_n}

\def\barRp{\barR_p}

\def\nmax{n_\textrm{max}}

\def\baR{\bar{\alpha}^{\scriptscriptstyle{\cal R}}}
\def\baRn{\baR_n}
\def\baRp{\baR_p}

\def\baQ{\bar{\alpha}^{\scriptscriptstyle{\cal Q}}}
\def\baQn{\baQ_n}
\def\baQp{\baQ_p}

\def\bbR{\bar{\beta}^{\scriptscriptstyle{\cal R}}}
\def\bbRn{\bbR_n}
\def\bbRp{\bbR_p}

\def\bQ{\beta^{\scriptscriptstyle{\cal Q}}}
\def\bQn{\bQ_n}

\def\bbQ{\bar{\beta}^{\scriptscriptstyle{\cal Q}}}
\def\bbQn{\bbQ_n}

\def\setsize{\csname @setfontsize\endcsname \setsize}

\begin{document}


\title{The CMB Bispectrum}

\author{J.R.~Fergusson}

\author{M.~Liguori}

\author{E.P.S.~Shellard}

\affiliation
{Centre for Theoretical Cosmology,\\
Department of Applied Mathematics and Theoretical Physics,\\
University of Cambridge,
Wilberforce Road, Cambridge CB3 0WA, United Kingdom}

\date{\today}

\abstract{
\noindent  We use a separable mode expansion estimator with WMAP7 data to 
estimate the bispectrum for all the primary families of non-Gaussian models, 
including non-scaling feature (periodic) models,
the flat (trans-Planckian) model, DBI  and ghost inflation, as well as previously constrained
simple cases.    We review the late-time mode expansion estimator methodology which can 
be applied to any non-separable primordial and CMB bispectrum model, and we 
demonstrate how the method can be used to reconstruct the CMB bispectrum from 
an observational map.   We extend the previous validation of the general estimator
using local map simulations.  We apply the estimator to the coadded WMAP 7-year V and W channel maps, 
reconstructing the WMAP bispectrum using $l<500$ multipoles and $n=50$ orthonormal 3D eigenmodes;
both the mode expansion parameters and the reconstructed 3D WMAP bispectrum are plotted. 
We constrain all popular nearly scale-invariant models, ensuring that the theoretical 
bispectrum is well-described by a convergent mode expansion.   Constraints from 
the local model $ \fnl=20.31 \pm 27.64$ and the equilateral model $\fnl=10.19 \pm 127.38$
($\Fnl = 1.90 \pm 23.79$) are consistent with previously published results.  (Here, 
we use a nonlinearity parameter $\Fnl$ normalised to the local case, to allow more direct 
comparison between different models.)  Notable new constraints from our method 
include those for the constant model $\Fnl = 7.82 \pm 24.57 $, the flat model 
$\Fnl = 7.31 \pm 26.22$, and warm inflation $\Fnl = 2.10 \pm 25.83$.   We investigate feature models,
which break scale invariance, surveying a wide parameter range for both the scale and phase (scanning for 
feature models with an effective period $l^*>150$).  We find no significant evidence of non-Gaussianity for all 
cases well-described by the given eigenmodes.  In the overall non-Gaussian 
analysis, we find one anomalous mode $n=33$ with a 3.39$\sigma$ amplitude 
which could give rise to an oscillatory model signal with $l^*\le 150$.  We propose a measure $\barFnl$ for the 
total integrated bispectrum and find that the measured value is consistent with the null 
hypothesis that CMB anisotropies obey Gaussian statistics.   We argue that this general bispectrum 
survey with the WMAP data represents the best test of Gaussianity to date and we discuss 
future prospects with higher precision and resolution, notably from the Planck satellite. 
}


\maketitle


\setsize{11}{13}

\section{Introduction}

\noindent In an earlier paper \cite{FergussonLiguoriShellard2009}, we described a general approach to 
the estimation of non-separable CMB bispectra using separable mode expansions.  
Our aim here is to directly estimate  the full CMB bispectrum from WMAP data, to survey 
and constrain current non-Gaussian primordial theories, and to 
discuss the prospects for reconstructing the bispectrum with forthcoming data, such as  the Planck 
experiment.    Previous work by other groups has endeavoured to measure the bispectrum by 
using specific estimators tailored to investigate particular separable models, such as the well-known  
local and equilateral bispectra.   This restriction to separable cases was for reasons of calculational 
simplicity to make the data analysis tractable, that is, reducing it from 
 ${\cal O}(\lmax^5)$ to  ${\cal O}(\lmax^3)$ operations. We summarise constraints 
that have been obtained to date using these methods later in section V, when we survey theoretical 
models; it is sufficient at this point to note that the present WMAP7 constraint  $-10<\fnl<74$ \cite{WMAP7}  
(95\% confidence) does not provide any significant evidence for a primordial local bispectrum signal, and nor do
constraints on the few  other models investigated to date (see the review \cite{LigSef2010}). 

Two significant developments mean that we can move beyond these specific estimators 
and consider a more general approach which includes the reconstruction of the whole bispectrum directly 
from the observational data.   First, explicit calculations of the reduced CMB bispectrum $\blll$  in a wide-ranging
survey of primordial theories \cite{Fergusson:2006pr, Fergusson:2008ra}, demonstrated that 
the resulting coherent patterns of acoustic peaks could be represented 
by rapidly convergent mode expansions with a limited number of terms (irrespective of 
whether the primordial bispectrum was separable).    
Secondly, these complete orthonormal mode expansions could be transformed 
into a non-orthogonal frame with separable basis functions \cite{FergussonLiguoriShellard2009} in which the 
same simplifications could be exploited to efficiently calculate the estimator (\ref{eq:optimalestimator})
in  ${\cal O}(\lmax^3)$ operations, again for arbitrary non-separable theoretical bispectra $\blll$.   
We shall employ this mode expansion methodology in this paper, convolving observational 
maps with the separable basis functions and then reconstructing the observed bispectrum $\blll$ 
in an expansion using the resulting mode coefficients.  
Rather than looking in just a few specific directions within the large space of 
possible bispectra, this general mode decomposition encompasses
all bispectra up to a given resolution.   Our aim is to determine whether there is evidence for
{\it any bispectrum} causing a departure from Gaussianity in the WMAP data.  Of course, 
we can compare with previous constraints for the local and equilateral models, but an important byproduct
is a set of entirely new constraints on a wide range of non-separable models.  

While we believe this work represents a significant step forward, we note that this analysis is far 
from the last word on  CMB non-Gaussianity, not least because much higher quality and higher 
resolution data will soon be available from Planck.   We also note that we have only used WMAP7 data out 
to l=500, together with a pseudo-optimal analysis of the noise and masking contributions.  
This paper should be considered primarily as a proof of concept implementation of these
methods, leaving up to an order of magnitude discovery potential available for bispectrum 
signals with new CMB data, let alone future galaxy and other 3D surveys where this approach
can also be applied.  We note that there are other recent methodologies in the literature which, 
in principle, can be used to extract information from the bispectrum beyond simple separable cases, 
including the bispectrum power approach of ref.~\cite{MunshiHeavens2009}), bispectrum 
binning used in ref.~\cite{BuchervanTent} and wavelet approaches (see the review \cite{LigSef2010}).

In section \ref{sec:CMBbispest} we review general results regarding primordial and angular 
bispectra and their optimal estimation. The eigenmode decomposition of the bispectrum that constitutes the 
foundation of our methodology is summarized in section \ref{sec:modeexp}. We then show in section 
\ref{sec:reconstruction} how this expansion can be used to reconstruct the full bispectrum from the data, 
before directly extracting the bispectrum from WMAP data in section \ref{sec:WMAP}. We then turn our attention to estimates of $\fnl$ for a wide 
variety of shapes, including both scale invariant bispectra (section \ref{sec:scaleinv}) and scale-dependent oscillatory 
bispectra (section \ref{sec:feature}), improved here using an inverse modal covariance \cite{Fergusson:2011sa}.. Finally, before drawing our conclusions in section \ref{sec:conclusions}, we discuss in section 
\ref{sec:totalbisp} a possible way to use our mode expansion technique to define 
a model independent constraint on the total integrated bispectrum extracted from the data.

\section{CMB bispectrum estimation}\label{sec:CMBbispest}

\subsection{Primordial and CMB bispectrum}

Temperature anisotropies are represented using the $a_{lm}$ coefficients of a spherical harmonic decomposition of 
the cosmic microwave sky, 
\eq \label{eq:alm}
\frac{\Delta T}{T}(\hat {\bf n}) = \sum_{lm} a_{lm} Y_{lm}(\hat {\bf n})\,,
\qe
with an (ideal) angular power spectrum $C_l = \sum _m a_{l\,m}\,a_{l\,-m}$.
The CMB bispectrum is the three point correlator of the $a_{lm}$, 
\eq\label{eq:fullbispectrum}
B^{l_1 l_2 l_3}_{m_1 m_2 m_3} = a_{l_1 m_1} a_{l_2 m_2} a_{l_3 m_3}\,,
\qe
where, here,  we assume that the $B^{l_1 l_2 l_3}_{m_1 m_2 m_3}$ coefficients are not an ensemble average but, instead, 
directly calculated using the $\alm$'s from a high resolution map (or maps), that is, from an
experiment such as WMAP or Planck.  
We shall assume for the moment that if there is a non-trivial bispectrum then it has arisen through a physical process which is statistically isotropic, so we can employ the angle-averaged bispectrum $\Blll$ without loss of information, that is \cite{Luo1994}, 
\eq
B_{l_1 l_2 l_3} &=& \sum_{m_i} \( \begin{array}{ccc} l_1 & l_2 & l_3 \\ m_1 & m_2 & m_3 \end{array} \)B^{l_1 l_2 l_3}_{m_1 m_2 m_3}\nn\\
&=&  \sum_{m_i}\hlll^{-1}  \curl{G}^{l_1 l_2 l_3}_{m_1 m_2 m_3} B^{l_1 l_2 l_3}_{m_1 m_2 m_3}\,,
\qe
where $\hlll$ is a geometrical factor,
\eq
\hlll =  \sqrt{\frac{(2l_1+1)(2l_2+1)(2l_3+1)}{4\pi}} \( \begin{array}{ccc} l_1 & l_2 & l_3 \\ 0 & 0 & 0 \end{array} \)\,,
\qe
and $ \curl{G}^{\,\,l_1\; l_2\; l_3}_{m_1 m_2 m_3}$ is the Gaunt integral,
\begin{align}\label{eq:Gaunt}
\nn \curl{G}^{l_1 l_2 l_3}_{m_1 m_2 m_3} &\equiv \int d\W \, Y_{l_1 m_1}(\un) \, Y_{l_2 m_2}(\un) \, Y_{l_3 m_3}(\un) \\
&=\hlll \( \begin{array}{ccc} l_1 & l_2 & l_3 \\ m_1 & m_2 & m_3 \end{array} \)\,,
\end{align}
with  the usual Wigner-$3j$ symbol ${\scriptstyle \big(\stackrel{\scriptstyle l_1 }{\scriptstyle m_1 }\stackrel{\scriptstyle l_2 }{\scriptstyle m_2 }\stackrel{\scriptstyle l_3 }{\scriptstyle m_3} \big)}$.   
It is more convenient to eliminate the geometrical factors entirely and to  work with the reduced bispectrum which is defined as 
\eq
\blll = \hlll^{-1} \Blll \,.
\qe

It is important to note the relationship between the late-time CMB bispectrum  $\blll$ and the 
primordial bispectrum $B_\Phi(\kall)$ from which it would arise in many models, notably inflation.
 The convention has been to remove a $k^{-6}$ scaling by defining a shape function:
\begin{align} \label{eq:shapefn}
S(k_1,k_2,k_3) \equiv \frac{1}{N} (k_1 k_2 k_3)^2 B_\O(\kall)\,.
\end{align}
The shape function (\ref{eq:shapefn}) is particularly pertinent for scale-invariant models because their momentum dependence is restricted entirely to planes transverse to the diagonal $\tilde k = {\textstyle {\frac{1}{2}}} (k_1+k_2+k_3)$.
The CMB bispectrum induced by the primordial shape $S$ is obtained from the convolution \cite{KomatsuSpergel2001}:
\begin{align} \label{eq:redbispect}
\nn b_{l_1 l_2 l_3}= \(\frac{2}{\pi}\)^3 \int x^2dx \int & d k_1 d k_2 d k_3\, S(k_1,k_2,k_3)\\
&\times  \D_{l_1}(k_1) \,\D_{l_2}(k_2)\, \D_{l_3}(k_3)\, j_{l_1}(k_1 x)\, j_{l_2}(k_2 x)\, j_{l_3}(k_3 x)\,,
\end{align}
where $ \D_{l}(k)$ is the transfer function.   

\begin{figure}[t]
\centering
\includegraphics[width=.48\linewidth]{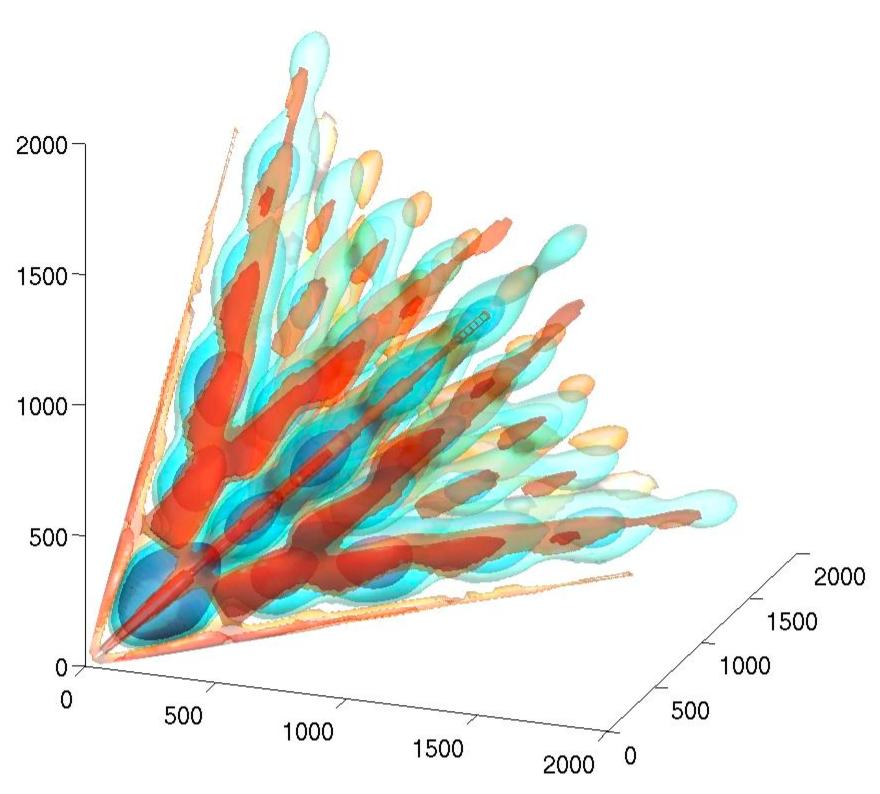}
\includegraphics[width=.48\linewidth]{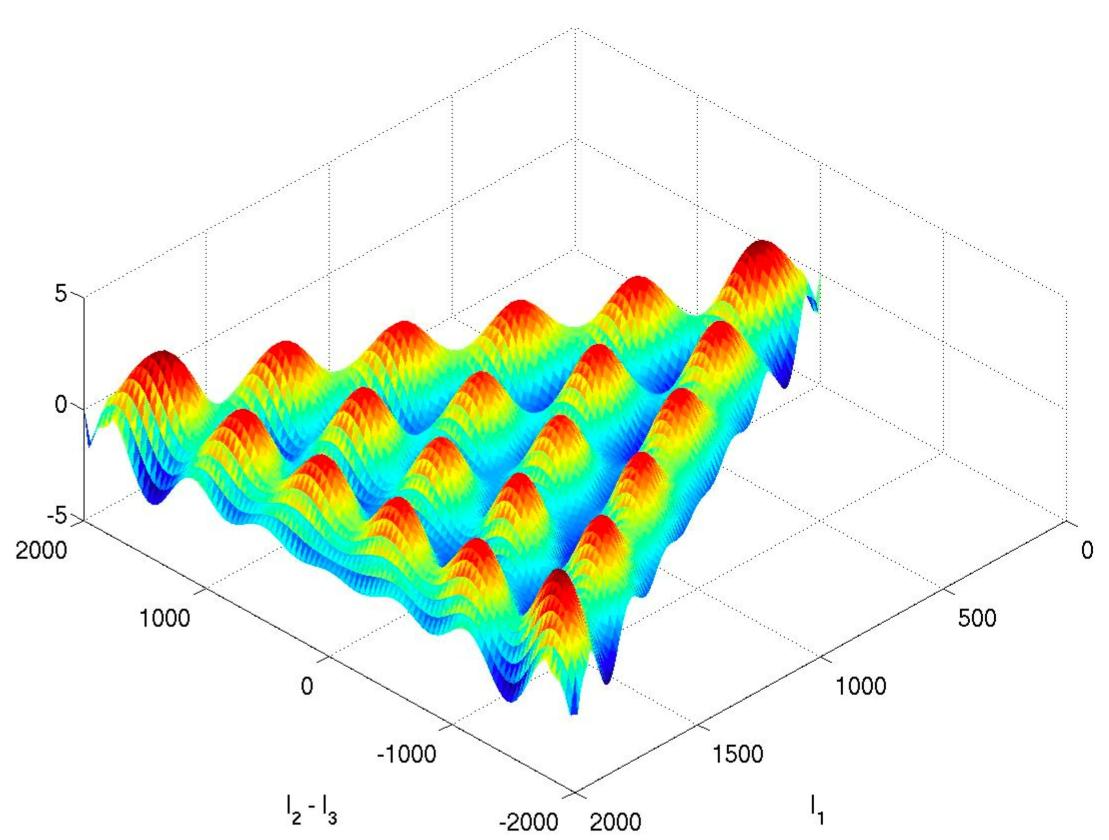}
\caption[Constant model]{\small The reduced CMB bispectrum  for the constant model $\blll^{\rm const}$ arising from the convolution of the primordial shape function $S(\kall) =1$
with transfer functions (normalised relative to the large-angle constant solution $\blll^{const(la)}$ given in (\ref{eq:constbispect})).
On the left,  the 3D bispectrum is plotted over the allowed tetrahedral region of multipole triples (see fig.~\ref{fig:tetrapyd}) using  several density contours  (light blue positive and magenta negative) out to 
$l_i\le 2000$.   On the right, a transverse triangular slices through the bispectrum is shown for  $\lsum= 4000$ (Planck
resolution).  
Note the coherent pattern of acoustic peaks with a dominant primary peak in a broad diagonal region around $l_1=l_2=l_3=220$. 
This constant model bispectrum plotted is the analogue of the angular power spectrum $C_l$'s for a purely scale-invariant model.  
 }
\label{fig:constant}
\end{figure}

The impact of the transfer functions  in (\ref{eq:redbispect}) is to impose 
a series of acoustic peaks on the underlying primordial shape, as illustrated for the CMB bispectrum of
 the constant model $S(\kall)=1$ in fig.~\ref{fig:constant}.    Here, we can observe a large primary 
peak when all the $l_i\approx 220$.   In principle, the CMB bispectrum is 
difficult to evaluate since (\ref{eq:redbispect}) represents a four-dimensional integral over highly 
oscillatory functions.   However, the integral breaks down into a product of one-dimensional integrals
if  the shape function is separable, 
that is, if it can be represented in the form $S(\kall) = X(k_1) Y(k_2)Z(k_3)$.  In the large-angle limit
with $\D_{l}(k) = j_l(...)$ ($l\ll 200$) it is possible in some separable models to obtain analytic solutions, such as that 
for the constant model  
\cite{Fergusson:2008ra}
\eq\label{eq:constbispect}
\blll^{\rm const(la)} = \frac{\D^2_\Phi}{27 N} \frac{1}{(2\ell_1+1)(2\ell_2+1)(2\ell_3+1)}\[\frac{1}{\ell_1+\ell_2+\ell_3+3} + \frac{1}{\ell_1+\ell_2+\ell_3}\]\,. 
\qe
This particular regular solution is important because we divide by it when plotting the CMB bispectrum $\blll/ \blll^{\rm const(la)}$ 
throughout this paper.  Normalising with the constant model (\ref{eq:constbispect}) is analogous to multiplying the 
power spectrum $C_l$'s by $l(l+1)$, because it serves to remove an overall  $\ell^{-4}$ scaling for all scale-invariant bispectra,  preserving the effects of the oscillating transfer functions without introducing spurious transverse momentum dependence.   This convention is quite similar to plotting the signal-to-noise $S/N$ (i.e.\ dividing by $(C_{l_1}C_{l_2}C_{l_3})^{1/2}$), except for the additional $l^{-1}$ term in square brackets in \eqref{eq:constbispect}, which effectively factors in the 2D CMB density of states $\propto l$.  Hence, division by $\blll^{\rm const(la)}$ allows us to democratically visualize bispectrum isosurfaces of the key regions which contribute to the estimator.

\subsection{CMB bispectrum estimators}

Now it is usually presumed that the full bispectrum for a high resolution map cannot be evaluated explicitly 
because of the sheer
number of operations involved ${\cal O}(\lmax^5)$, as well as the fact that the signal will be too weak to measure 
individual multipoles with any significance.    Instead, we essentially use a least squares fit to compare the bispectrum 
of the observed $\alm$'s (\ref{eq:alm}) with a particular (separable) theoretical bispectrum $\blll^{\rm th}$, 
\eq
\langle a^{\rm th}_{l_1 m_1} a^{\rm th}_{l_2 m_2} a^{\rm th}_{l_3 m_3}\rangle=  \curl{G}^{\,\,l_1\; l_2\; l_3}_{m_1 m_2 m_3 }
\blll^{\rm th}\,.
\qe 
Here, $\blll^{\rm th}$ will be recovered as the expectation value from an ensemble average over  $\alm^{\rm th}$ realisations 
or simulations created with the given reduced bispectrum.  
Formally,  taking into account the fact that instrument noise and masking can break rotational invariance, the result is the general optimal estimator 
\cite{KSW,CreminellietAl2006,SmithetAl2009}
\eq\label{eq:optimalestimator}
{\curl{E}} = \frac{1}{{{N}^2}} \sum_{l_i,m_i}& &\left[  \curl{G}^{\,\,l_1\; l_2\; l_3}_{m_1 m_2 m_3 } \blll ^{\rm th}\(C^{-1}_{l_1 m_1, l_4 m_4} a_{l_1m_1}\) 
                   \(C^{-1}_{l_2 m_2, l_5 m_5} a_{l_2m_2}\) \(C^{-1}_{l_3 m_3, l_6 m_6}a_{l_3m_3} \) \right.
                   \nonumber \\
            & &   ~~~ \left. -~3 \left \langle a_{l_1m_1} a_{l_2m_2} a_{l_3m_3} \right \rangle C^{-1}_{l_1 m_1, l_2 m_2} 
                   C^{-1}_{l_3 m_3,l_4 m_4} a_{l_4m_4} \right]\,,
 \qe
where $C^{-1}$ is the inverse of the covariance matrix $C_{\ell_1 m_1, \ell_2 m_2} = \langle a_{\ell_1 m_1}a_{l_2 m_2}\rangle$
and  $\ {N}$ is  a suitable normalisation (discussed further below).   Here, we  follow ref.~\cite{WMAP5,YadavWandelt2009}, by assuming a 
nearly diagonal covariance matrix ($C_{l_1 m_1, l_2 m_2} \approx C_l \,\delta_{l_1l_2}\,\delta_{m_1\,-m_2}$) and approximating 
the estimator (\ref{eq:optimalestimator}) as
\begin{align}\label{eq:approxestimator}
\curl{E} = \frac{1}{\tilde{N}^2} \sum_{l_i m_i} \frac{\curl{G}^{l_1 l_2 l_3}_{m_1 m_2 m_3} \, \tilde{b}_{l_1 l_2 l_3} }{ \tilde{C}_{l_1}\tilde{C}_{l_2}\tilde{C}_{l_3} } \(a_{l_1 m_1} a_{l_2 m_2} a_{l_3 m_3}  - 6\, C^{\rm sim}_{l_1 m_1 , l_2 m_2}  a_{l_3 m_3}\)\,,
\end{align}
where the tilde denotes the modification of  $C_l$ and $\blll$ to incorporate instrument beam and noise effects through  
\begin{align}
\label{eq:noisebeam}
\tilde{C}_l = b_l^2 C_l + N_l   \qquad \mbox{and}\qquad \tilde{b}_{l_1 l_2 l_3} = b_{l_1}b_{l_2}b_{l_3}\, b_{l_1 l_2 l_3}\,.
\end{align}
For a relatively small galactic mask (leaving a large fraction $f_{\rm sky}$ of the full sky), it has also been shown to be a good 
approximation to renormalise using  
\eq
\label{eq:cutsky}
\blll^{\rm mask} = f_{\rm sky} \blll  \qquad \mbox{and}\qquad C_l^{\rm mask} =  f_{\rm sky} C_l\,.
\qe
(We shall assume noise,  beam and mask inclusion henceforth and drop any special notation.)
Here, the second linear term  in (\ref{eq:approxestimator}) ensures subtraction of spurious inhomogeneous noise 
and masking contributions by using the covariance matrix $C^{\rm sim}_{l_1 m_1 , l_2 m_2}$ from an  
ensemble average of Gaussian maps in which these effects are incorporated. 

If the theoretical bispectrum 
$\blll^{\rm th}$ has the property of primordial separability then it has been noted that the summation in (\ref{eq:approxestimator}) 
becomes much more tractable taking only  ${\cal O}(\lmax^3)$ operations \cite{KSW}.  Essentially this
exploits the separability of the Gaunt integral (\ref{eq:Gaunt}), as well as primordial counterparts, to 
reduce the dimensionality of the integrals and summations involved in evaluating (\ref{eq:approxestimator}) 
(see ref.~\cite{FergussonLiguoriShellard2009} for a more detailed discussion on this point). To date, such separability has been 
a property of all the primordial theories constrained observationally with most attention given to the canonical local model.   

\subsection{$\Fnl$ normalisation}

It remains to briefly discuss the normalisation factor $N$ in (\ref{eq:optimalestimator}).   In the past this has been 
taken on a case-by-case manner for a given theoretical bispectrum $\blll^{\rm th}$  to be 
\eq\label{eq:theorynorm}
{N_{\rm th}}^2 \equiv \sum_{l_i} \frac{\hlll^2{b^{\rm th}_{l_1 l_2 l_3}}^2}{C_{l_1}C_{l_2}C_{l_3}}\,.
\qe 
As we discuss below, this has yielded very model-dependent results for the measurement of the 
nonlinearity parameter $\fnl^{\rm th} \equiv {\cal E}$.   Instead, we have proposed the parameter
$\Fnl$ which is much easier to compare between models, because it measures the 
integrated CMB bispectrum signal relative to that from the canonical local model with $\fnl=1$.   In this case, we define\cite{FergussonLiguoriShellard2009}
\eq \label{eq:newfnl}
\Fnl^{\rm th} = {\cal E}, \quad\mbox{with}\quad N^2 \equiv N_{\rm loc}N_{\rm th} \,,
\qe
with $N_{\rm th}$ from (\ref{eq:theorynorm}) and where $N_{\rm loc}$ is defined for the $\fnl=1$ local model:
\eq 
\quad {N_{\rm loc}}^2  \equiv  \sum_{l_i} \frac{\hlll^2{b^{{\rm loc}(\fnl=1)}_{l_1 l_2 l_3}}^2}{C_{l_1}C_{l_2}C_{l_3}}\,.
\qe
Of course, for the local model, the quantities are identical $\Fnl^{\rm th} = \fnl^{\rm loc}$. 
However, when we quote constraints on other models we will use $\Fnl$---making self-evident the comparable nature 
of this quantity---while also noting the $\fnl^{\rm th}$ previously used in the literature.

The problem with $\fnl^{\rm th}$ is that it derives from a somewhat arbitrary normalisation 
of the primordial bispectrum $B^{\rm th}_\Phi(k_1,k_2,k_3)$ which bears little relation to the 
observable CMB bispectrum signal.   The convention has been to assume 
 a nearly scale-invariant shape function $S(\kall)$ and then to normalise it such that $S^{\rm th}(k,k,k)=1$, that is, 
 at a single point; this  becomes $\fnl=1$ case for the model under study.   This definition
ignores the behaviour away from the equilateral value $k$$=$$k_1$$=$$k_2$$=$$k_3$.  For example, $S$  rises 
from a central minimum in 
the local model and falls from a maximum in the equilateral model; hence, the huge disparities between their $\fnl$ 
constraints, e.g. $\Delta \fnl^{\rm equil} \approx 7 \Delta \fnllocal$.   This definition also does not apply to non-scaling 
models.  The alternative to base the non-Gaussianity measure on the actually observable CMB bispectrum $\blll^{\rm th}$, as above in (\ref{eq:newfnl}), does accommodate non-scale invariant models, such as feature models.  It also covers bispectra 
induced by late-time processes like gravitational lensing and cosmic strings.  For models which are 
not scale-invariant it should be quoted with the observational cut-off $\lmax$.   The normalisation for a particular model $\Fnl^{\rm th}$ can be easily forecast using the primordial $B_\Phi^{\rm th}(\kall)$ without the need for accurate CMB calculations of $\blll^{\rm th}$ in (\ref{eq:redbispect}); primordial shape autocorrelators just need to be compared with the local shape as 
demonstrated in ref.~\cite{Fergusson:2008ra}.

\section{Separable mode expansions}\label{sec:modeexp}

\begin{figure}[t]
\centering
\includegraphics[width=.6\linewidth]{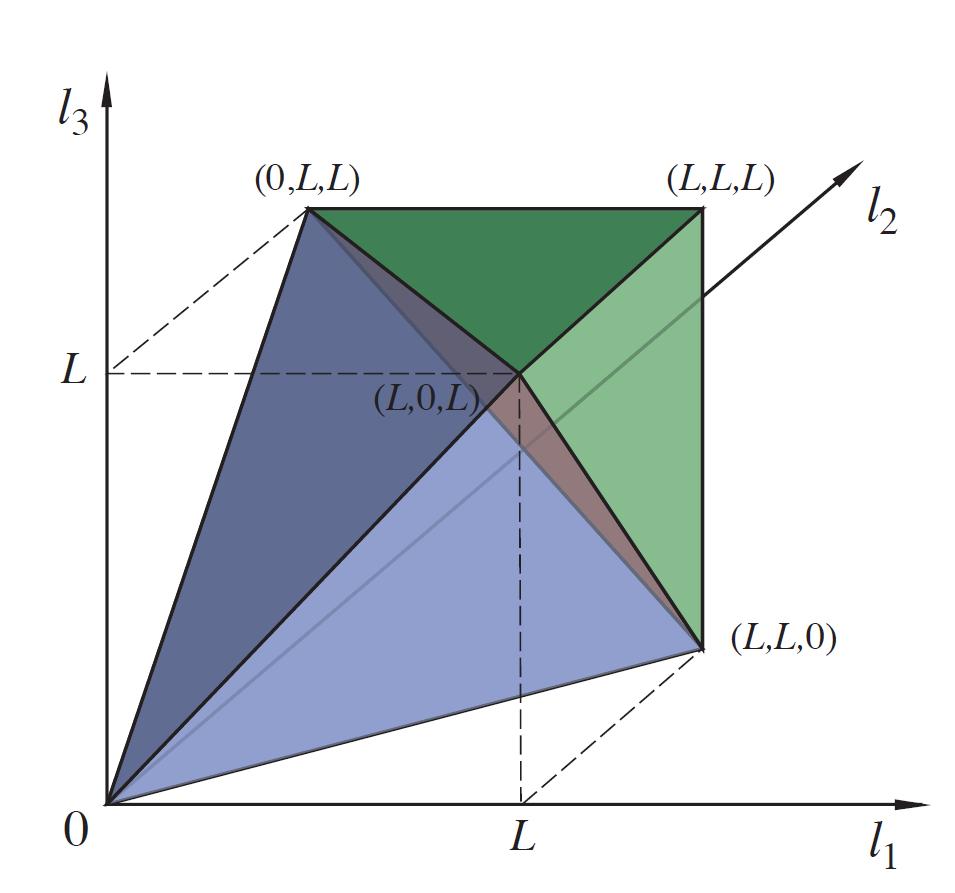}
\caption[Tetrahedral domain]{\small  Observational domain (\ref{eq:tetrapydl}) for the CMB bispectrum $\blll$.  Allowed multipole values $(l_1,\,l_2,\,l_3)$ lie inside the shaded  `tetrapyd' region, satisfying both the triangle condition and $l <L $$\,\equiv\,$$\lmax$.}
\label{fig:tetrapyd}
\end{figure}

When analysing the CMB bispectrum $\blll$, we are restricted  to a tetrahedral domain of 
multipole triples $\{l_1l_2l_3\}$ satisfying both a triangle condition and a limit given by the maximum resolution 
$\lmax$ of the experiment.  This three-dimensional domain $\Vtetra$ of allowed multipoles is illustrated in 
fig.~\ref{fig:tetrapyd} and it is explicitly defined by 
\eq \label{eq:tetrapydl}
\nonumber
&&\mbox{Resolution:} \qquad \qquad~ \lall \leq \lmax\,,\quad \lall \in  \mathbb{N}\,,\\
&&\mbox{Triangle condition:}\quad l_1 \leq l_2+l_3  ~~\hbox{for}~~ l_1 \geq l_2,\,l_3,  ~~  +~\hbox{cyclic}~\hbox{perms.}\,,\\ \nonumber&&\mbox{Parity condition:} \qquad\lsum = 2n\, ,~~~n\in\mathbb{N}\,.
\qe
The multipole domain is denoted a `tetrapyd'
because it arises from the union of a regular tetrahedron from the origin out to the plane $\lsum\le 2\lmax$ and a triangular pyramid constructed from the corner of the cube taking in the remaining multipole values out to $l_i\le \lmax$.    Summed bispectrum expressions such as (\ref{eq:theorynorm}) indicate that we must define a weight function $w_{l_1 l_2 l_3}$ on the tetrapyd domain
in terms of the geometrical factor $\hlll$, that is, 
\eq\label{eq:lweightdiscrete}
w_{l_1 l_2 l_3} =\hlll^2\,.
\qe
This is a nearly constant function on cross sections defined by $l_1+l_2+l_3=\mbox{const}$, except very near the tetrahedral boundaries where it is still bounded, and a useful and accurate continuum limit $w(\lall)$ is given in \cite{FergussonLiguoriShellard2009}.
In order to eliminate an $l^{-1/2}$ scaling in the bispectrum estimator functions, we usually 
exploit the freedom to divide by a separable function and to employ instead the weight 
\eq  \label{eq:lweightsep}
w_s(\lall) = \frac{w_{l_1 l_2 l_3} }{v_{l_1}^2v_{l_2}^2v_{l_3}^2}\,,\quad \mbox{where} \quad v_l = (2l+1)^{1/6}\,.
\qe
We can then define an inner product of two functions $f(\lall),\,g(\lall)$ on the tetrapyd domain (\ref{eq:tetrapydl})
through 
\eq\label{eq:innerproduct}
\langle f,\,g\rangle ~\equiv~ \sum_{\lall\in\Vtetra } w_s(\lall)\, f(\lall)\, g(\lall)\,.
\qe
Given that calculations generally deal with smooth functions $f,\,g,\,w,\, v$, we can use a variety of schemes to speed up 
this summation (effectively an integration).

Our goal is to represent the observed CMB bispectrum estimator functions, such as those in (\ref{eq:approxestimator}) and (\ref{eq:theorynorm}), on the multipole domain (\ref{eq:tetrapydl})
using a separable mode expansion, 
\eq \label{eq:cmbestmodes}
\frac{v_{l_1}v_{l_2}v_{l_3}}{\sqrt{C_{l_1}C_{l_2}C_{l_3}}} \, \blll = \sum_n \baQn \barQn(\lall)\,,
\qe 
where the $\barQn$ are basis functions constructed from symmetrised polynomial products 
\eq
\barQn (\lall) &=& {\textstyle \frac{1}{6}}[\bar q_p(l_1)\, \bar q_r(l_2)\, \bar q_s(l_3) +  \bar q_r(l_1)\, \bar q_p(l_2)\, \bar q_s(l_3) +  \mbox{cyclic perms in $prs$}]\nn\\
&\equiv& \bar q_{\{p} q_{r}q_{s\}}\quad \mbox{with}\quad n\leftrightarrow \{prs\}\,,
\qe
with the $\bar q_p(l)$ defined below.   Here, the six permutations of the polynomial products which we denote as $\{ prs\}$ 
reflect the underlying symmetries of the bispectrum $\blll$ . For convenience, we define a one-to-one mapping  $n\leftrightarrow \{prs\}$ ordering the permuted triple indices into a single list labelled by $n\in \mathbb{N}$.    Alternative `slicing' and `distance' orderings were presented in  ref.~\cite{FergussonLiguoriShellard2009}, but the results presented here are robust to this change.   However, we shall quote explicit coefficients $\bQn$ resulting from distance ordering (i.e.\ $n(\lall) < n'(l_1',l_2',l_3')$ implies $l_1^2+l_2^2+l_3^2\le{l_1'}^2+{l_2'}^2+{l_3'}^2$ and in the instance of two modes being equidistant the one with most equal $l_i$ takes precedence).

We choose to define the tetrahedral $\bar q_p(l)$ polynomials analogously to 
Legendre polynomials $P_n$ by requiring them to be self-orthogonal with respect to the 
inner product (\ref{eq:innerproduct}),
\eq
\langle\bar q_p(l_1),\,\bar q_r(l_1)\rangle = \delta_{pr}\,,
\qe
with the first few polynomials given by $\bar q_0=0.074$,  $\bar q_1 = 0.30(-0.61+l)$, $\bar q_2 = 1.2(0.26 -1.1\,l +l^2)$ etc. 
More precise expressions and generating functions are given in ref.~\cite{FergussonLiguoriShellard2009}.
As products, the $q_p$  only confer partial orthogonality on the 3D basis functions $\barQn$, 
but their use is vital  for other reasons, given their bounded and near scale-invariant behaviour.

While the product basis functions $\barQn$ are independent and separable, they are not orthogonal in general
\eq
\langle \barQn,\,\barQp\rangle \equiv \gamma_{np}\ne \delta_{np}\,,
\qe
so it is very useful to construct a related set of orthonormal mode functions $\barRn$ using Gram-Schmidt orthogonalisation 
such that
\eq\label{eq:orthonormal}
\langle  \barRn,\,\barRp\rangle = \delta_{np}\,.
\qe
Working up to a given order $N$, the two sets of  mode functions are related through 
\begin{align}\label{eq:RQinverse}
\curl{R}_n = \sum_{p=0}^n \l_{mp} \curl{Q}_p \quad \hbox{for}~~ n,p\le N\,,
\end{align}
where $ \l_{mp}$ is a lower triangular matrix with 
\eq\label{eq:gammalambda}         
(\l^{-1})_{np} ^{\top} = \langle\Qn,\,\Rp\rangle \qquad\mbox{and} \qquad  
(\gamma^{-1})_{np} = \sum_{r}^N(\l^\top)_{nr}\l_{rp}\, .
\qe
Knowing $\l_{np}$ allows us to easily systematically evaluate
the expansion coefficients in (\ref{eq:cmbestmodes}) directly from the inner product 
\eq \label{eq:RQrelation}
\baRn =\Big \langle \barRn,\, \frac{v_{l_1}v_{l_2}v_{l_3}}{\sqrt{C_{l_1}C_{l_2}C_{l_3}}} \, \blll\Big\rangle\,,
~~~~\hbox{yielding}~~~~\baQn = \sum_{p=0}^{N}(\l^\top)_{np}\, \baRp\,.
\qe
Indeed, it is more convenient to present our final bispectrum results in the orthonormal $\barRn$ basis,
\eq \label{eq:cmborthmodes}
\frac{v_{l_1}v_{l_2}v_{l_3}}{\sqrt{C_{l_1}C_{l_2}C_{l_3}}} \, \blll = \sum_n \baRn \barRn\,
\qe 
 because their orthonormality 
(\ref{eq:orthonormal}) implies a version of Parseval's theorem.  Here, we note that the expansion (\ref{eq:cmborthmodes})
presumes a spectrum normalised as in (\ref{eq:newfnl}) to have $\Fnl =1$, that is, with $N$ such that $\sum_n\baRn{}^2=N^2$
in the estimator (\ref{eq:approxestimator}). 

To summarise, the $\barQn(l_1,l_2,l_3)$'s are independent separable 
basis functions built out of the permutations of simple products of the polynomials $\bar q_p(l)$, which 
are well-behaved and bounded over the tetrapyd.  The  $\barQn$'s in their easily separable form
are employed directly in the bispectrum estimator.   However, it is more straightforward to present results and to 
use the inner product (\ref{eq:innerproduct}) with the transformed $\barRn$ eigenmodes because they are orthonormal; a simple 
matrix expression (\ref{eq:RQrelation}) relates the expansion coefficients $\baQn$ and $\baRn$ using the two sets of basis functions. 

\section{Reconstructing the CMB bispectrum}\label{sec:reconstruction}

Now consider the implications of substituting the mode expansion (\ref{eq:cmbestmodes}) into the estimator (\ref{eq:approxestimator}), while exploiting the separability of the Gaunt integral (\ref{eq:Gaunt}),
\eq\label{eq:cmbestsep}
{\cal E} &=& \frac{1}{N^2}\sum_{l_i,m_i}\sum _{n\leftrightarrow prs}\kern-6pt \baQn\bar q_{\{p}\bar q_r \bar q_{s\}} \int d^2\hat {\bf n} \frac{Y_{l_2m_2}(\hat {\bf n})Y_{l_1m_1}(\hat {\bf n})\, Y_{l_3m_3} (\hat {\bf n})}{{v_{l_1}v_{l_2}v_{l_3}}\sqrt{C_{l_1}C_{l_2}C_{l_3}}}
\left[a_{l_1m_1}a_{l_2m_2}a_{l_3m_3} - 6 \langle a_{l_1m_1}a_{l_2m_2}\rangle a_{l_3m_3}\right]\nn\\
&=& \frac{1}{N^2} \sum _{n\leftrightarrow prs}\kern-6pt \baQn  \int d^2\hat {\bf n}\left[\(\sum_{l_1,m_1} \bar q_{\{p} \, \frac{a_{l_1m_1} Y_{l_1m_1}}{v_{l_1} 
\sqrt{C_{l_1}}}\)\( \sum_{l_2,m_2} 
\bar q_{r} \, \frac{a_{l_2m_2} Y_{l_2m_2}}{v_{l_2} \sqrt{C_{l_2}}}\)\(
 \sum_{l_3,m_3} \bar q_{s\}} \, \frac{a_{l_3m_3} Y_{l_3m_3}}{v_{l_3} 
\sqrt{C_{l_3}}}\)\right.\nn\\
&&~~~~~~-6\left.\left\langle\(\sum_{l_1,m_1} \bar q_{\{p} \, \frac{a_{l_1m_1} Y_{l_1m_1}}{v_{l_1} 
\sqrt{C_{l_1}}}\)\( \sum_{l_2,m_2} \bar q_{r} \, \frac{a_{l_2m_2} Y_{l_2m_2}}{v_{l_2} \sqrt{C_{l_2}}}\)\right\rangle\(
 \sum_{l_3,m_3} \bar q_{s\}} \, \frac{a_{l_3m_3} Y_{l_3m_3}}{v_{l_3} 
\sqrt{C_{l_3}}}\)\right] \\
&=& \frac{1}{N^2}  \sum _{n\leftrightarrow prs}\kern-6pt \baQn \int d^2\hat {\bf n}\,\left[\bar M_{\{p}(\un)\bar M_r(\un)\bar M_{s\}}(\un)-
6\left\langle\bar M^{\rm G}_{\{p}(\un)\bar M^{\rm G}_r(\un)\right\rangle\bar M_{s\}}(\un)\right]\,.
\qe 
Here, the $\bar M_p(\un)$ represent versions of the original CMB map filtered with 
the polynomial $\bar q_p$ with the separated weight function $(v_l \sqrt{C_l})^{-1}$, that is, 
\begin{align}\label{eq:barmapfilter}
\bar M_p(\un) = \sum_{lm} \bar q_p(l)\frac{a_{lm}}{v_l\sqrt{C_l}} Y_{lm}(\un)\,.
\end{align}
The maps  $\bar M^{\rm G}_p(\un)$ incorporate the same mask 
and a realistic model of the inhomogeneous instrument noise; a large ensemble of these maps, calculated from Gaussian simulations, are used in the averaged linear
term in the estimator (\ref{eq:cmbestsep}), allowing for the subtraction of these important effects.  Defining the integral over
these convolved product maps as cubic and linear terms respectively 
\eq\label{eq:mapintegral}   
\bbQn{}^{\rm cub} &=&  \int d^2\hat {\bf n}\, \bar M_{\{p}(\un)\bar M_r(\un)\bar M_{s\}}(\un)\,,\\
\bbQn{}^{\rm lin} &=&  \int d^2\hat {\bf n}\, \left\langle\bar M^{\rm G}_{\{p}(\un)\bar M^{\rm G}_r(\un)\right\rangle\bar M_{s\}}(\un)\nn\,,
\qe
the estimator (\ref{eq:approxestimator}) reduces to a simple sum over the mode coefficients
\eq
{\cal E} = \frac{1}{N^2} \sum_n \baQn \bbQn\,,
\qe
where $\bbQn \equiv  \bbQn{}^{\rm cub} - \bbQn{}^{\rm lin}$.
This can be improved by introducing the inverse covariance of the the beta (which is best calculated in the orthonormal space, $\mathcal{R}$) so the estimator becomes \cite{Fergusson:2011sa}
\eq\label{eq:estimatorsum}
{\cal E} = \frac{1}{N^2} \sum_{np} \baRn \z^{-1}_{np} \bbRn \,,
\qe   
where $\z_{np} = \< \bbRn \bbRp \>$ is the covariance of $\bbRn$ and we have modified the normalisation similarly, $N = \sum_{np} \baRn \z^{-1}_{np} \baRp$. This addition will reduce the variance by taking into account the correlations in $\bbRn$, improving the optimality of the result.
The modal estimator can approach optimality by incorporating additional anisotropic modes accounting for the effect of the mask and then inverting the covariance \eqref{eq:estimatorsum} (see a recent publication for a more general discussion \cite{Fergusson:2011sa}).

The estimator sum (\ref{eq:estimatorsum})  is straightforward to evaluate, provided
the theoretical model coefficients $\baQn$ are known.  It has  been separated into
a product of three sums over the observational maps (\ref{eq:cmbestsep}), followed by a single integral over
all directions (\ref{eq:mapintegral}).   The actual operations entailed in the estimator sum are only ${\cal{O}} (l^2)$,
so these late-time methods are extremely rapid for direct data analysis and for obtaining variances from map
simulations.   However, we note that the preparatory `one-off' calculations setting up the orthonormal eigenmodes
and theoretical CMB bispectra are of order ${\cal{O}} (l^3)$.   We emphasise that the 
utility of this approach depends on a fairly rapidly convergent expansion for the theoretical bispectrum under 
study (as indicated for almost all models studied to date \cite{Fergusson:2008ra}) and the fact that we have 
constructed a {\it complete} set of orthonormal eigenmodes on the observed multipole domain (\ref{eq:tetrapydl}). 

There is potentially much more information in the observed $\bbQn$ coefficients than just the estimator
sum (\ref{eq:estimatorsum}) which only yields $\fnl$ for a given theoretical model.  Following the steps above in 
(\ref{eq:cmbestsep}), it is easy to show (see Appendix)
that the expectation value for $\bbQn$ for an ensemble of maps with a given CMB bispectrum (expanded in modes $\baRn$
with amplitude $\Fnl$) is  
\eq
\langle \bbQn\rangle = \sum_p\Fnl \baQn \langle \barQn,\, \barQp\rangle =\Fnl \sum_p\baQn\gamma_{np}\,,
\qe
so that the averaged estimator (\ref{eq:estimatorsum}) becomes 
\eq\label{eq:estimatorsum2}
\langle{\cal E}\rangle = \frac{1}{N^2} \Fnl\sum_n\sum_p \baQn\, \gamma_{np}\, \baQp  = \frac{1}{N^2} \Fnl\sum_n \baRn{}^2 = \Fnl\,,
\qe
where we have used (\ref{eq:RQrelation}) in transforming to the $\barRn$ basis. (Here we note again that in this basis 
$N^2 = \sum_n \baRn{}^2$.)
Equivalently, then, in the orthonormal frame we have the simple result  
\eq\label{eq:bestfitbeta}
\langle \bbRn\rangle =\Fnl \baRn\,,
\qe 
that is, we expect the best fit $\bbRn$ coefficients for a particular realization to be the $\baRn$'s themselves (given
a sufficiently large signal).  
Assuming that we can extract the $\bbRn$ coefficients with sufficient significance from a particular experiment, 
this means that we can directly reconstruct the CMB bispectrum using the expansion (\ref{eq:cmborthmodes}).

\section{The WMAP bispectrum}\label{sec:WMAP}

\begin{figure}[b]
\centering
\includegraphics[width=.9\linewidth, height = 6.5cm]{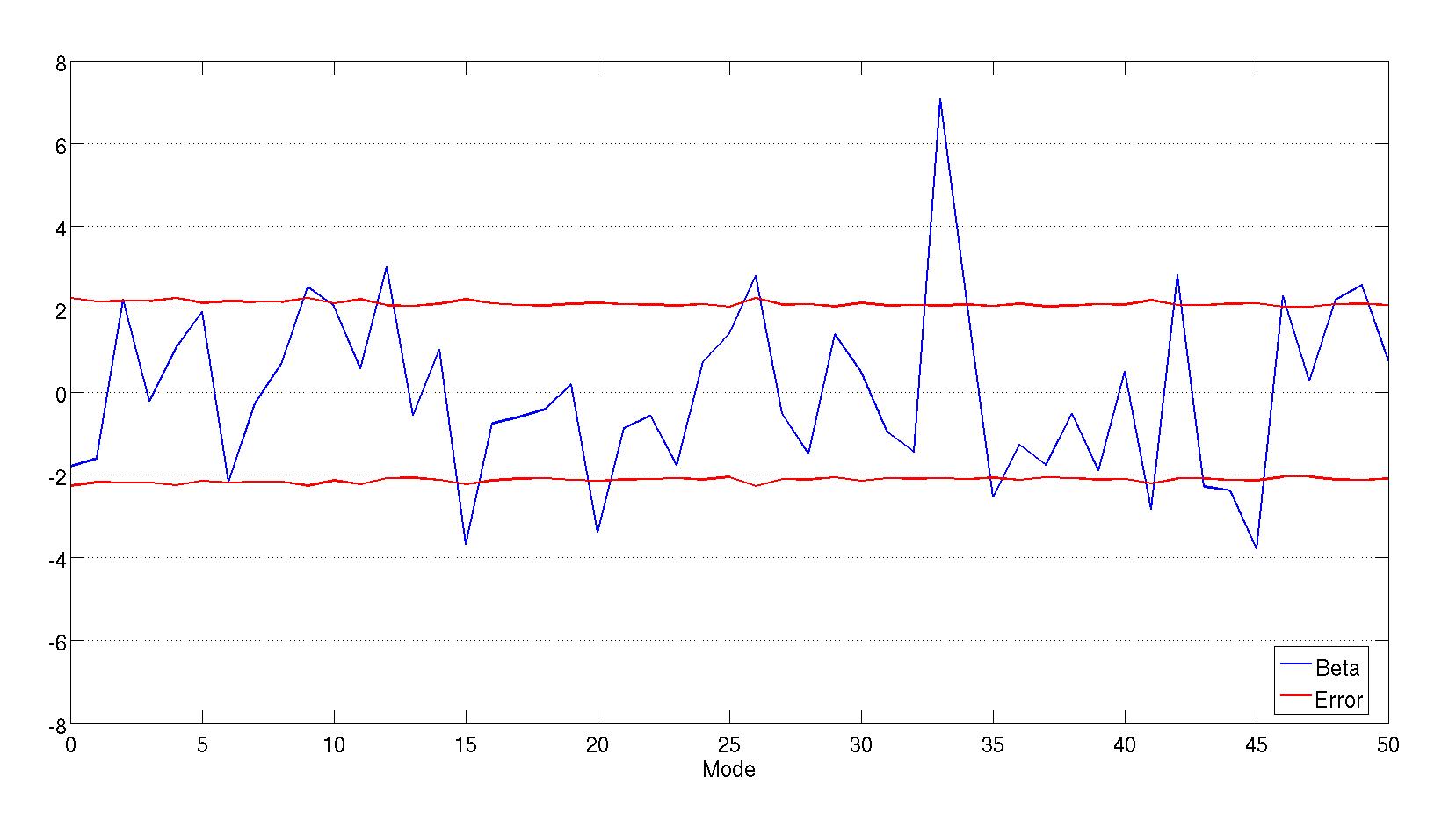}
\caption[]{\small  Recovered mode coefficients $\bbRn$ (\ref{eq:cmbestmodes}) from 
the WMAP7 coadded V and W maps.   Error bars (1$\sigma$) are also shown 
for each mode as estimated from 144000 Gaussian map simulations in WMAP-realistic context.  }
\label{fig:reconalphaWMAP5}
\end{figure}

\begin{figure}[t]
\centering 
\includegraphics[width=.85\linewidth]{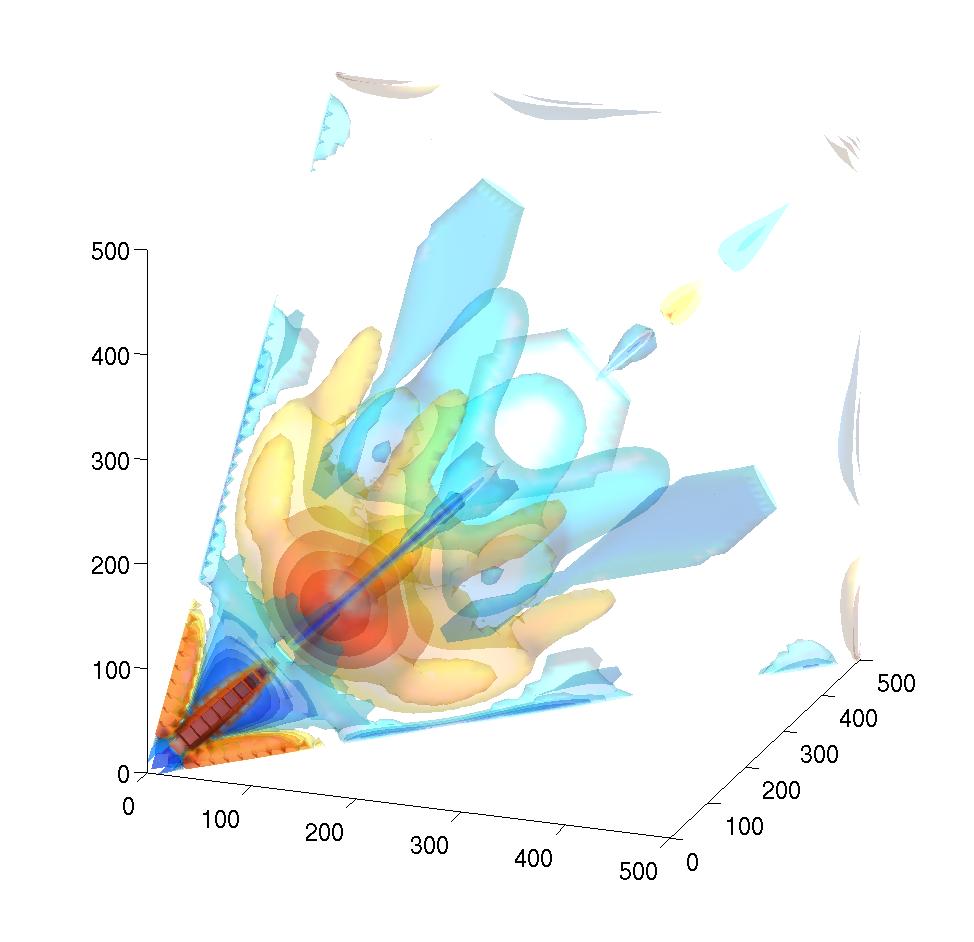}
\caption[Recovered bispectrum]{\small  Recovered 3D bispectrum from WMAP7 data showing the result
using the reconstructed mode coefficients $\bbRn$ shown in fig.~\ref{fig:reconalphaWMAP5} with the 
partial sum (\ref{eq:cmborthmodes}).  Several isodensity surfaces are shown for the bispectrum  out to 
$l_i\le 500$ (light blue positive and magenta negative). }
\label{fig:3dreconWMAP5}
\end{figure}

\begin{figure}[t]
\centering
\includegraphics[width=.825\linewidth]{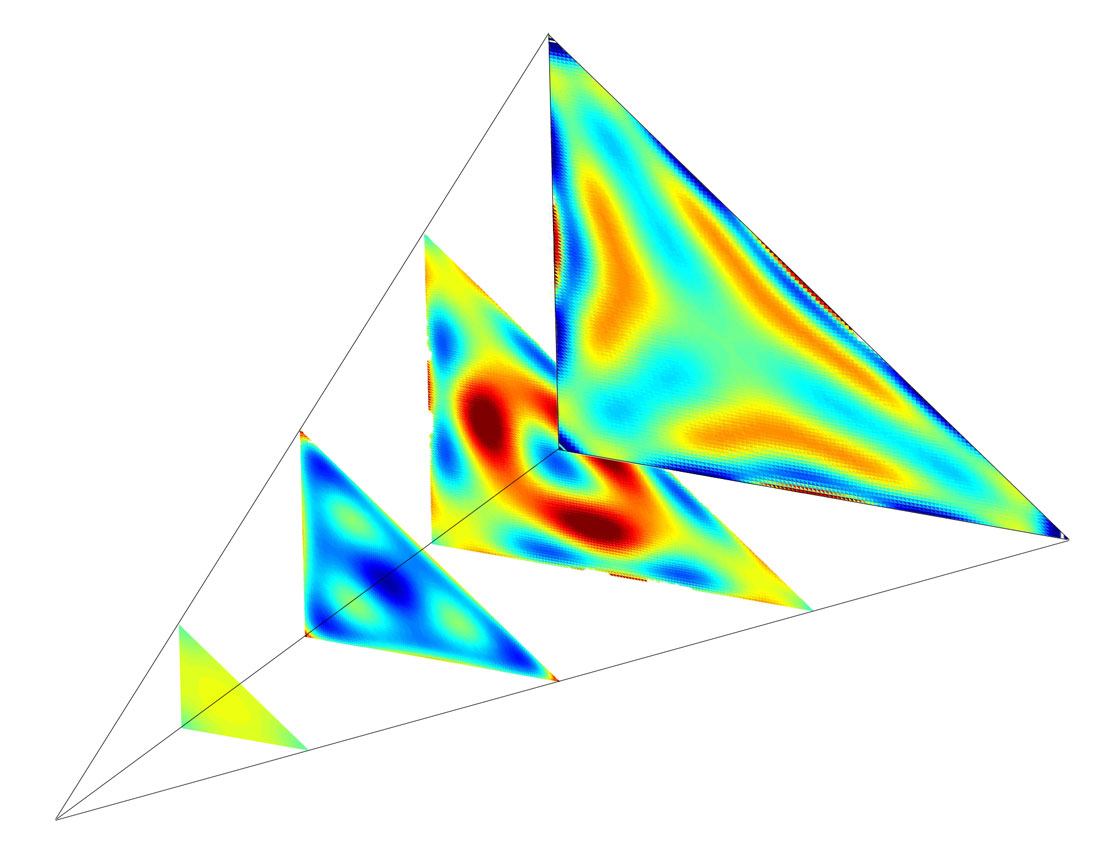}
\caption[Recovered bispectrum]{\small   
Recovered 3D bispectrum from WMAP7 data showing slices through the data 
at $\tilde l\equiv l_1+l_2+l_3= \hbox{const.}$.  Slices shown are $\tilde l = 250, 500, 750, 1000$, using the 
same colour scale as fig.~\ref{fig:3dreconWMAP5}.}
\label{fig:3dreconslices}
\end{figure}

We now move on to apply the mode decomposition techniques described and validated 
in the previous sections to the analysis of WMAP7 data. Our aim, first, will be
to estimate $\fnl$ arising from different primordial shapes, some as yet unconstrained in the literature 
(such as the feature models of section \ref{sec:nonscaling} and the flattened models of section \ref{sec:flattened}).
Secondly, we aim 
to provide a  full reconstruction of the bispectrum from the data, using the same pipeline shown to recover
local and equilateral bispectra from simulated data. 
The main emphasis of this work is obtaining 
fast and accurate convergence for many different shapes, rather than a fully optimised estimation. 
The analysis presented here is intended as a proof-of-concept for 
late time modal estimators of non-Gaussianity, gleaning valuable new information from WMAP rather achieving
a maximal extraction. For this reason our study has a number of limitations, which we enumerate here. We do not implement full inverse covariance weighting in the estimator 
as in (\ref{eq:optimalestimator}) \cite{SmithetAl2009}, but we adopt the pseudo-optimal weighting scheme used by 
the WMAP team for the WMAP 5-year analysis 
\cite{WMAP5}, improved by the inverse modal covariance  \eqref{eq:estimatorsum}; we 
use multipoles up to $\ell_{\max} = 500$, rather than $1000$, since the pseudo-optimal $\fnl$ error bars tend to 
saturate above that threshold; finally, we work with WMAP 7-year substantially revising our earlier WMAP 5-year data results. However, we note that the WMAP 5-year data was originally studied with a pseudo-optimal weighting approach, so comparison 
between our results and previous work was straightforward and showed consistency.   The present work represents the initial implementation of this general approach to analysing non-Gaussianity, rather than its completion even for the WMAP data given that we have used $\lmax<1000$. 
 
After coadding the V and W band data (with the same weights as in the WMAP7 
analysis), our first step was to extract the $\bbQn$ mode coefficients from the data, following the procedure summarized eqn (\ref{eq:barmapfilter}) and (\ref{eq:mapintegral}). In our analysis we chose to compute the first $n=51$ modes in (\ref{eq:estimatorsum}) because this proved sufficient to describe almost all theoretical CMB bispectra on the observational
domain $\lmax=500$. The resulting estimates 
will be shown in the following sections.  As pointed out in (\ref{eq:bestfitbeta}), by rotating 
our recovered $\bbQn$ into the orthonormal frame we obtain the best-fit estimate of the actual bispectrum coefficients $\baRn$.
The mode coefficients obtained from the WMAP7 data $\bbRn$ in this orthonormal frame are  plotted in 
fig.~\ref{fig:reconalphaWMAP5}.   The variance and $\zeta$ are estimated from 144,000 Gaussian map simulations, using the pipeline
repetitively in the same WMAP-realistic context. 

The mode coefficient extraction from the WMAP7 data was straightforward with both the cubic and 
linear terms contributing significantly to the final result.    The late-time estimator (\ref{eq:approxestimator})
is sensitive to all forms of non-Gaussianity, in contrast to the two or three separable (and 
oscillating)  modes previously extracted from the data using primordial estimators.  Despite this increased sensitivity, in principle,
making the method more susceptible to foreground contamination, 
our results do not appear to have been significantly affected after subtraction by the linear term.   
This has been investigated through extensive testing, including increasing mask size, and we will discuss these issues at much 
greater length in a companion paper \cite{FergLigShell2012},  characterising the mask, noise and other contributions.  It is interesting 
to note here, however, that the mode decompositions also can be used to characterise spurious anisotropic contributions, such as the inhomogeneous noise (and other contaminants).  We will show quantitatively how the action of the linear term 
essentially projects out these spurious bispectrum directions from the cubic term in  (\ref{eq:approxestimator}).  The local shape is the most affected
(as noted originally in ref.~\cite{CreminellietAl2006}).    

We note that in fig.~\ref{fig:reconalphaWMAP5} there is one anomalous $3.4\sigma$ mode at $n=33$ with this choice of polynomical basis and ordering.    The existence of an anomalous signal on the apparent lengthscale probed by this mode was also exhibited with an alternative basis and ordering.  While this is interesting and may signify some primordial signal or some instrumental or foreground effect, as we shall see, this mode does not correlate well with the theoretical models under study in this paper.  Of course, statistically we should expect to see some $2\sigma$$+$ mode coefficients, given that there are $\nmax=51$ of them.  Nevertheless, the apparent height of the $n=33$ mode is accentuated because the WMAP7 signal is generally lower than expected given the simple experimental noise model employed (see last section).  We are investigating the robustness of this signal using the full WMAP7 data ($\lmax =1000$) with alternative trigonometric basis functions at higher resolution \cite{FergLigShell2012}.  

The extracted mode coefficients $\bbRn$ from  
fig.~\ref{fig:reconalphaWMAP5} can be used to reconstruct the full 3D WMAP bispectrum using (\ref{eq:cmborthmodes}).   
The result of this partial sum is shown in fig.~\ref{fig:3dreconWMAP5}, together with a series of transverse slices through 
the bispectrum shown in  fig.~\ref{fig:3dreconslices}.   Visually the WMAP bispectrum bears some  qualitative resemblance
to the local CMB bispectrum in the squeezed limit which could  be consistent with some local or local-type contribution, but the periodicity of the 
other features does not match well with scale-invariant primordial models (whose periodicities are determined entirely by the transfer functions).   
The orthonormal mode coefficients $\bbRn$ plotted in fig.~\ref{fig:reconalphaWMAP5} do not individually show significant deviations
away from Gaussianity (but for one anomaly), given the nearly constant mode variances which are also plotted.   We note at the outset, therefore, 
that the WMAP bispectrum shown in fig.~\ref{fig:3dreconWMAP5} is likely to be the result of cosmic variance 
(perhaps with some residual local signal left-over from the noise/mask subtraction and or other contamination).   
As well as constraining specific theoretical models, we shall test the assumption of Gaussianity  more generally in 
section \ref{sec:totalbisp}   by considering a measure of the total integrated bispectrum 
obtained from the squared coefficients $\bbRn{}^2$.  In the near future, using the full WMAP7 data set and smaller variances we 
will expand the scope of our mode exploration, including principal component analysis and other statistical approaches \cite{FergLigShell2012}.   

Before obtaining specific new constraints, we emphasise again that the extraction of the mode coefficients $\bbRn$ provides a completely model-independent assessment of the three-point correlation function.  The approach provides far more 
information  than that contained in a simple $\fnl$ amplitude parameter extraction for particular models.   
Although obvious deviations from Gaussianity
are not apparent from this limited WMAP7 analysis (i.e.\ pseudo-optimal error bars and $\lmax=500$), there remains considerable 
potential with new data sets.  For Planck, the sensitivity to primordial non-Gaussianity will improve by 
up to an order of magnitude and so the error bars in   fig.~\ref{fig:reconalphaWMAP5} will shrink dramatically.   The prospects
for detection of a large NG signal remain completely open.

\section{Constraints on nearly scale-invariant models}\label{sec:scaleinv}

Constraints on the bispectrum to date have been for scale-invariant models of separable form,
primarily on the local and equilateral models, discussed 
previously.   There has been significant evolution over time for these constraints as both the CMB 
data and the estimation methodology have improved.   However, as table~\ref{tab:review}
illustrates (taken from ref.~\cite{LigSef2010}), there is no compelling and confirmed evidence for 
a significant non-Gaussian signal at this stage.   Our purpose in this section is to apply our 
more general mode expansion estimator (\ref{eq:estimatorsum}) with our WMAP analysis to obtain constraints 
on a much wider set of scale-invariant models.  This method can be applied to any model for 
which there is good convergence with the given $\nmax$ modes.     

\begin{table}[h]
\begin{tabular}{c |cl | cl }
       & \multicolumn{2}{c}{\bf Local} & \multicolumn{2}{c}{\bf Equilateral} \\ 
       \hline
   \multirow{2}{*}{\bf Pure cubic} & $-58 < \fnl < 134$ &WMAP1\cite{WMAP1} & $-366 < \fnl < 238$ & WMAP1 \cite{CreminellietAl2006}  \\ &  $-54 < \fnl < 114$ & WMAP3 \cite{WMAP3} &  $-256 < \fnl <332$  &WMAP3 \cite{CreminellietAl2006} 
   \\
  \hline 
  \multirow{5}{*}{\bf Pseudo-optimal}  & $-27 < \fnl < 121$ & WMAP1 \cite{CreminellietAl2006}  &  $-151 < \fnl < 253$ & WMAP5 \cite{WMAP5}  \\ & $-36 < \fnl < 100$  & WMAP3 \cite{CreminellietAl2006} &  \\  & $\,\,27\, < \fnl < \, 147$ & WMAP3 \cite{YadavWandelt2009}   &  \\ & $9\,\, < \fnl < 129\,$ & WMAP3\cite{SmithetAl2009} &
   \\ & $\,-9\, < \fnl < \, 111$ &  WMAP5 \cite{WMAP5} &  \\
  \hline
\multirow{3}{*}{\bf Optimal} & $12<\fnl<104$ & WMAP3 \cite{SmithetAl2009}  & $-125<\fnl<435$& WMAP5 \cite{SmithetAl2009}  \\ & $-4<\fnl<80$ & WMAP5 \cite{SmithetAl2009}  & $-254<\fnl<306$& WMAP7 \cite{WMAP7}        \\
& $-10<\fnl<74$ & WMAP7 \cite{WMAP7}  &        \\
   \hline
\end{tabular}
\caption{Constraints on $\fnl^{local}$,$\fnl^{equil.}$, obtained by different groups on the one-year (W1), three-year (W3), five-year (W5), and seven-year (W7) 
WMAP data releases. The estimators employed are the pseudo-optimal (\ref{eq:approxestimator}), the cubic (the same 
without the linear noise term), and the optimal with full-covariance weighting (\ref{eq:optimalestimator}).  All results were in 
the context of a primordial estimator using separable functions to describe the specific model, unlike the general 
late-time estimator employed here.  For further details about the estimator methods employed and 
the significant evolution of these results over time, please refer to the review \cite{LigSef2010}. }
\label{tab:review}
\end{table}

\begin{figure}[b]
\centering
\includegraphics[width=.75\linewidth]{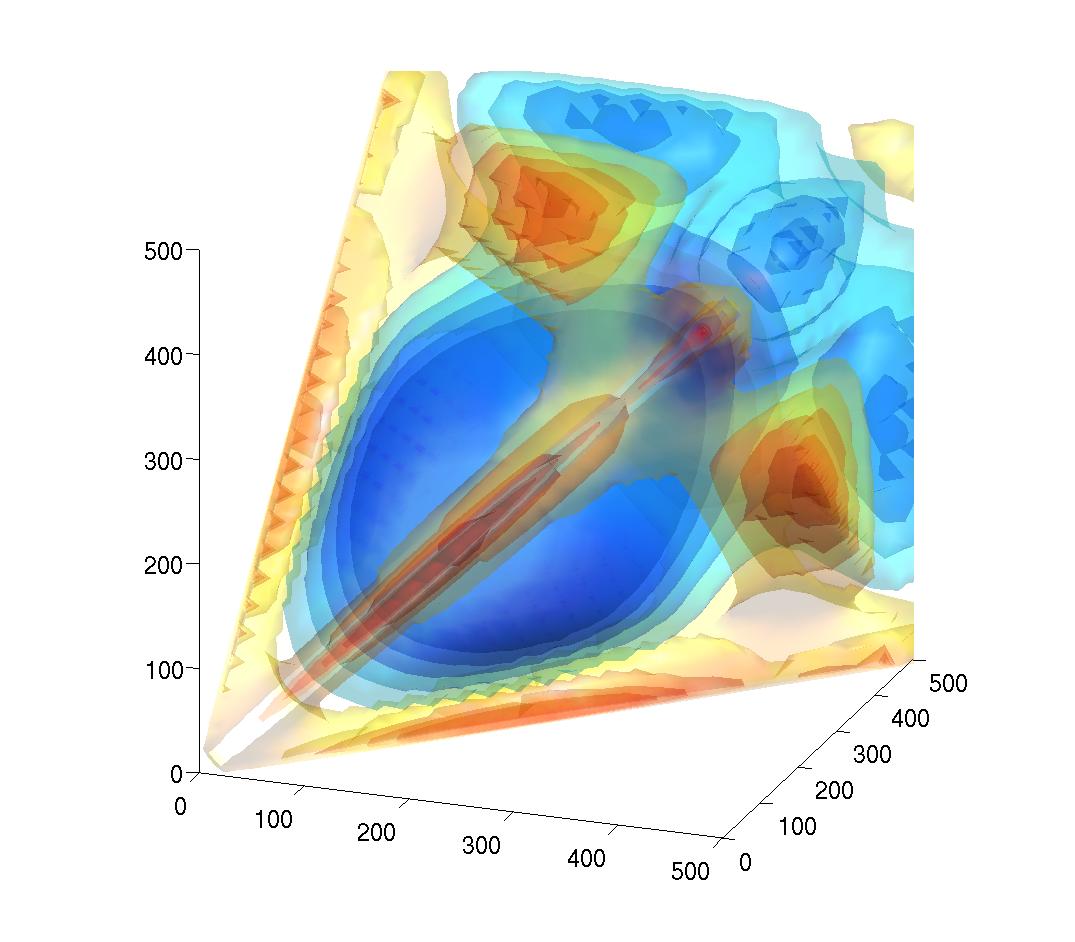}
\caption[Recovered bispectrum]{\small  
Predicted 3D bispectrum for the constant model up to $l_i\le 500$. 
 The same thresholds are employed as those shown in the 
WMAP reconstructions in fig.~\ref{fig:3dreconWMAP5} 
(after an overall rescaling). }
\label{fig:const3Dslices}
\end{figure}

\subsection{Constant model}

The constant model $S(\kall)= 1$ is the simplest possible primordial shape with triangles of every configuration 
contributing equally, resulting in a CMB bispectrum $\blll$ with features due entirely to the transfer functions (as we observed
for the acoustic peaks shown in fig.~\ref{fig:constant}). The constant model was motivated initially 
by its simplicity with the large-angle analytic solution (\ref{eq:constbispect}) for the CMB bispectrum \cite{Fergusson:2008ra}.
However, the constant shape does have other more explicit physical motivation, such as 
generation during a slowly turning trajectory in multifield inflation, denoted quasi-single field inflation \citep{ChenWang2009}.  
For nearly scale-invariant models, the central values for the bispectrum, $b_{lll}$, all have roughly the same profile but with different normalisations. The oscillatory properties of the transfer functions create acoustic peaks located at triple combinations involving the following  multipole values, 
 $l \approx 200, 500, 800, ...$.    To observe the key differences between scale invariant models we must study the bispectrum in the plane orthogonal to the $(l,l,l)$-direction, that is, the directions reflecting changes in the primordial shape functions.   
 
 For the multipole
 range $\lmax <500$ relevant to the present analysis, we have plotted the 3D bispectrum in fig.~\ref{fig:const3Dslices}. Here, the dominant feature is the primary acoustic peak stretched along the diagonal 
 of the tetrapyd, peaking at $l = l_1=l_2=l_3=220$ and elongated like an extended balloon from $l\approx100$ to $l\approx450$.  
 Evidence for this primary peak would indicate the presence of a primordial and scale-invariant non-Gaussian signal, as 
 emphasised in ref.~\cite{Fergusson:2008ra} and investigated quantitatively for the local model in  ref.~\cite{BuchervanTent}.
 Observing the reconstructed WMAP bispectrum shown in fig.~\ref{fig:3dreconWMAP5} there is a central fluctuation at $l\approx 140$
 but it does not extend to larger $l$ as would be expected; see the $l_1+l_2+l_3 =750$ slice in  fig.~\ref{fig:3dreconslices} (right)
corresponding to $l\approx 250$ where the (apparent) WMAP peak has disappeared.   If this measured 3D WMAP bispectrum were considered
to have any statistical significance then it would mitigate against a scale-invariant model, motivating the discussion in section~\ref{sec:nonscaling}.
 
\begin{figure}[t]
\centering
\includegraphics[width=.9\linewidth,height=5cm]{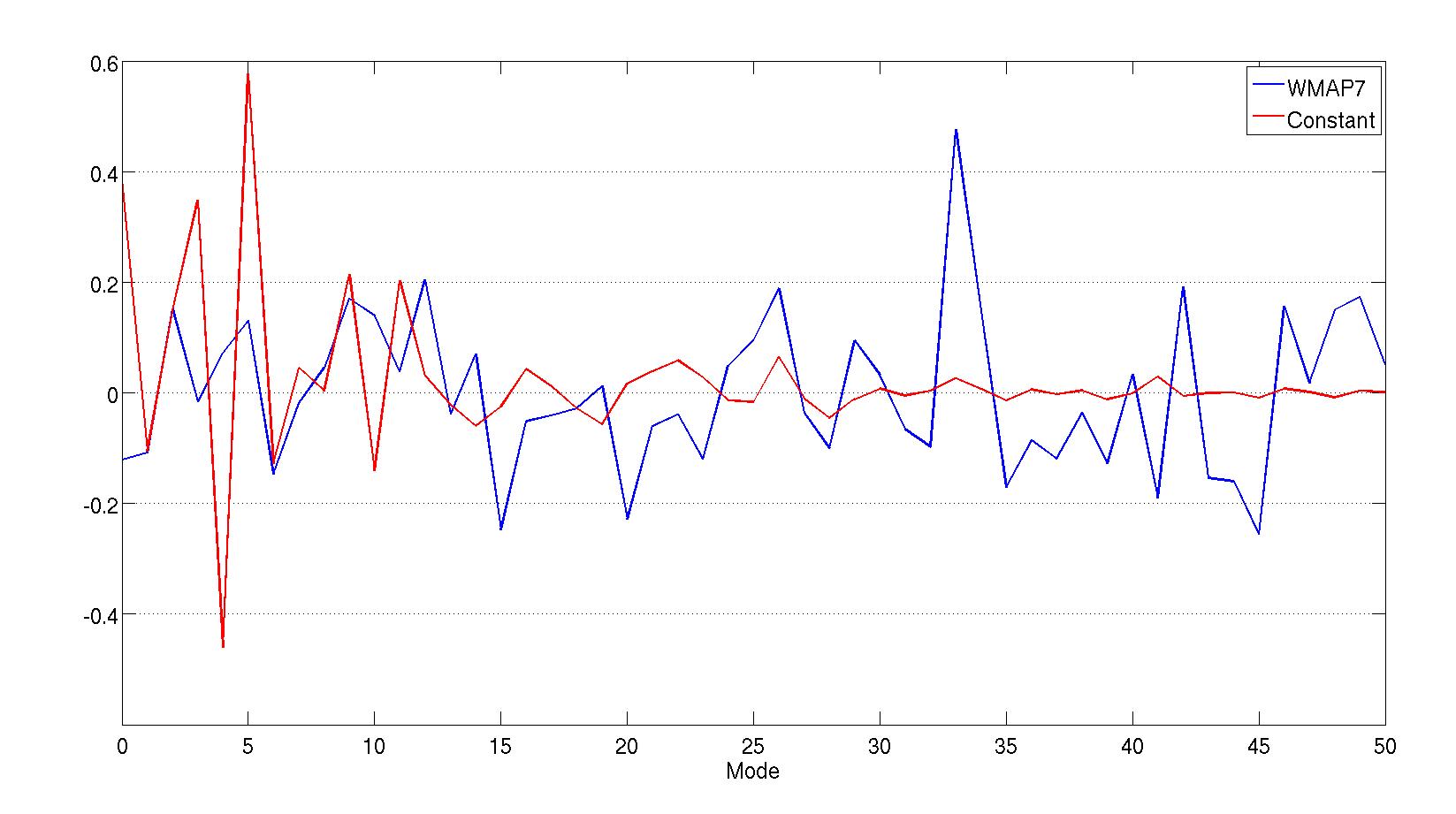}
\caption[]{\small Comparison between constant model and 
recovered mode coefficients for the WMAP7 data.   Note that the constant model incorporates features
entirely due to the transfer functions (the acoustic peaks seen in modes $n=3,4,5$), which are indicators of 
its primordial origin. }
\label{fig:constalpha}
\end{figure}

A comparison of the mode coefficients $ \baRn{}^{\rm const} $ from the constant model CMB bispectrum shown in fig.~\ref{fig:constalpha} indicates little obvious correlation with the WMAP coefficients $ \bbRn{}^{\rm wmap} $ (also plotted).  
Note that the constant model mode coefficients are large for the constant offset $n=0$ and for $n=3,4,5$ reflecting the periodicity of the acoustic peak structure (for $\lmax=500$), that is,  corresponding to the $\bar q_p\bar q_r\bar q_s$ polynomial products
with  $prs = \{000\}, \{002\}, \{111\}, \{012\}$ 
(also with related harmonics at lower amplitude with $n=9,10,11$).
The mode decomposition estimator (\ref{eq:estimatorsum}) yields the quantitative 
constraint 
\eq
\label{eq:fnlconst}
\Fnl^{\rm const} = 7.82 \pm 24.57 \,,\qquad \quad(\fnl^{\rm const}= 30.53 \pm 95.92)\,,
\qe
where $\Fnl$ is the bispectrum parameter normalised relative to the local model (\ref{eq:newfnl}), 
while the lower case $\fnl$ constraint 
employs the more model-dependent normalisation using the primordial shape function $S(k,k,k)=1$.  
It is clear from this result that there is no evidence---given
the present precision---for a significant constant primordial non-Gaussian signal.

\subsection{Local model}

The canonical local shape  covers a wide range of models where the non-Gaussianity is produced by local interactions. These models have their peak signal in ``squeezed" states where one $k_i$ is much smaller than the other two due to non-Gaussianity typically being produced on superhorizon scales. Single-field slow-roll inflation is dominated by the local shape, though $\fnllocal$ is tiny \cite{Maldacena2003,AcquavivaetAl2003}. The production of large 
non-Gaussianity during multiple field inflation \cite{RigopoulosShellardvanTent2006A,SeeryLidsey2005,VernizziWands2006} shows much greater promise of producing an observable signal through conversion of isocurvature  into adiabatic perturbations.   
Large  $\fnllocal$ can also be produced in curvaton models \citep{LindeMukhanov2006, LythUngarelliWands2003, BartoloMatarreseRiotto2004}, at the end of inflation from reheating mechanisms \cite{EnqvistetAl2005A} and also in more exotic scenarios such as (non-local) $p$-adic inflation \cite{BarnabyCline2008} and  the ekpyrotic scenario \cite{LehnersRenauxPetel}.   
For more comprehensive references and recent examples please refer to the review, ref.~\cite{Chen2010}.

\begin{figure}[t]
\centering
\includegraphics[width=.9\linewidth, height = 5cm]{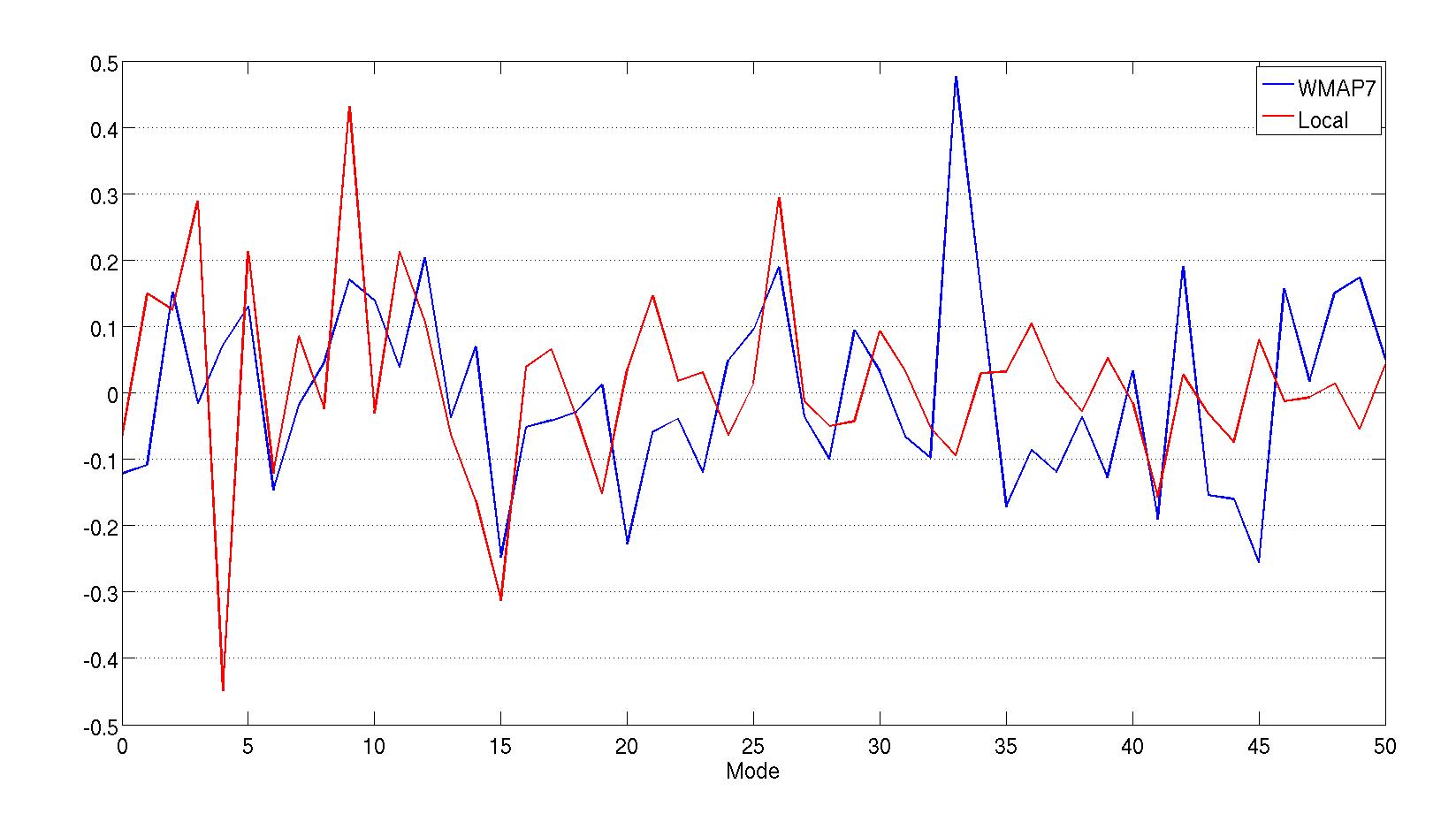}
\caption[]{\small  Comparison between local model expansion coefficients and 
recovered modes for the WMAP7 data.   Note the relatively slow convergence
of the local model and the apparent visual correlation of modes. }
\label{fig:localalpha}
\end{figure}

The distinct mode decomposition of the local model is illustrated in fig.~\ref{fig:localalpha}, together with the WMAP7 spectrum.  
The local model expansion is quite distinct  from the 
constant model reflecting the dominant signal along the edges of the tetrahedron, and favouring the higher order
polynomials needed to describe this localised signal.  That is, as well as the periodic acoustic peak signal seen 
in the constant model ($n=3,4,5$), the spectrum is otherwise
dominated by pure modes $n=9,\,15,\,26$ with $prs = \{003\}, \{004\}, \{005\}$.  The expansion is not as rapidly convergent but
the eigenmode partial sum achieves a 98\% correlation by $n=51$.

To aid comparison with the recovered WMAP bispectrum, we illustrate both in fig.~\ref{fig:localalpha}.   There appears to be 
some  correlation between the two sets of data points which is reflected in the result from the mode estimator 
\eq
\label{eq:fnllocal}
\Fnllocal = 20.31 \pm 27.64 \qquad (\fnllocal=20.31 \pm 27.64)\,.
\qe

This result is consistent, but slightly smaller than, that found by other groups. The small difference could easily be explained by the use of different $l_{max}$ in the other analyses.

\subsection{Equilateral models}

\begin{figure}[b]
\centering
\includegraphics[width=0.32\linewidth,height=0.25\linewidth]{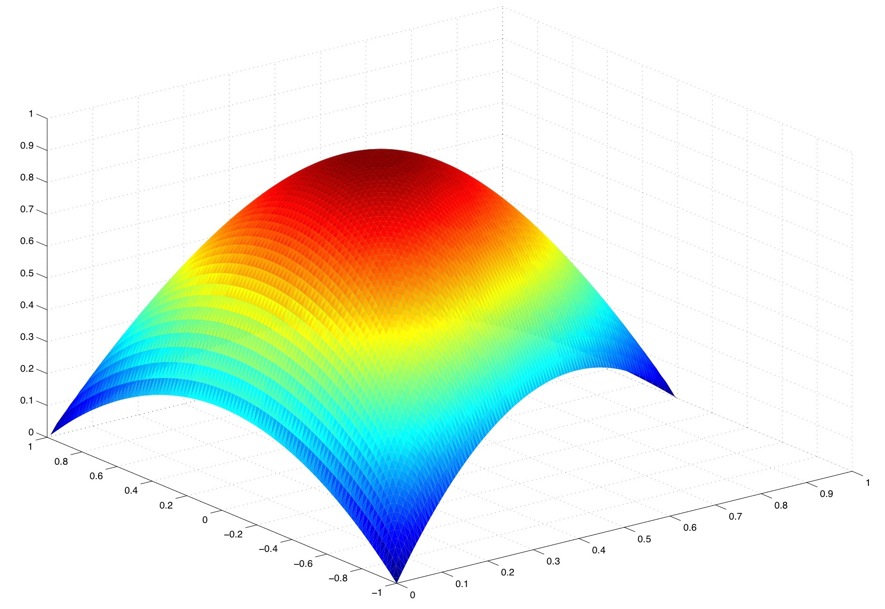} 
\includegraphics[width=0.32\linewidth,height=0.25\linewidth]{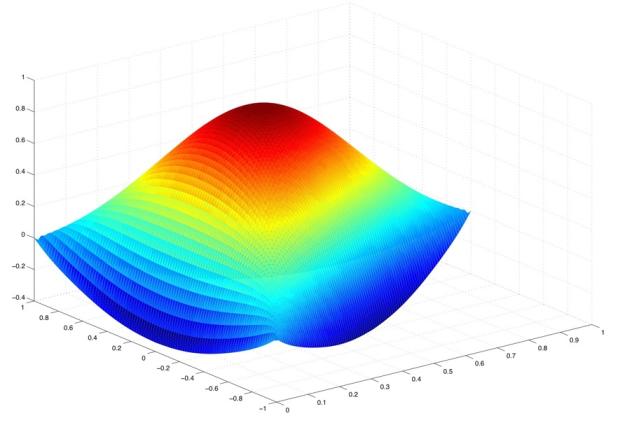}
\includegraphics[width=0.32\linewidth,height=0.25\linewidth]{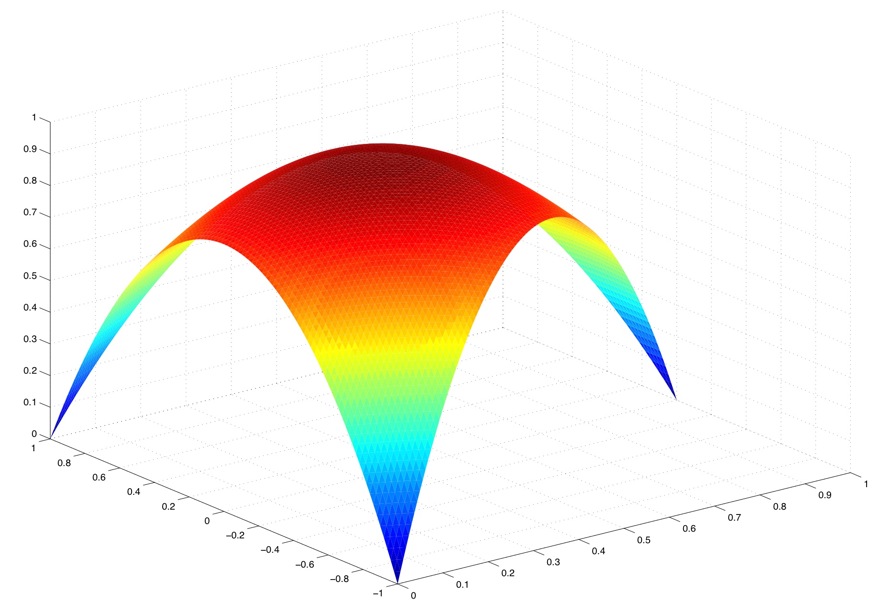}
\caption[The shape function of models in the equilateral class.]{\small The shape function of models in the equilateral class
which from left to right are DBI inflation,  ghost inflation  and the remaining single field inflation model.}
\label{fig:equipics}
\end{figure}

\begin{figure}[b]
\centering
\includegraphics[width=.9\linewidth, height = 5cm]{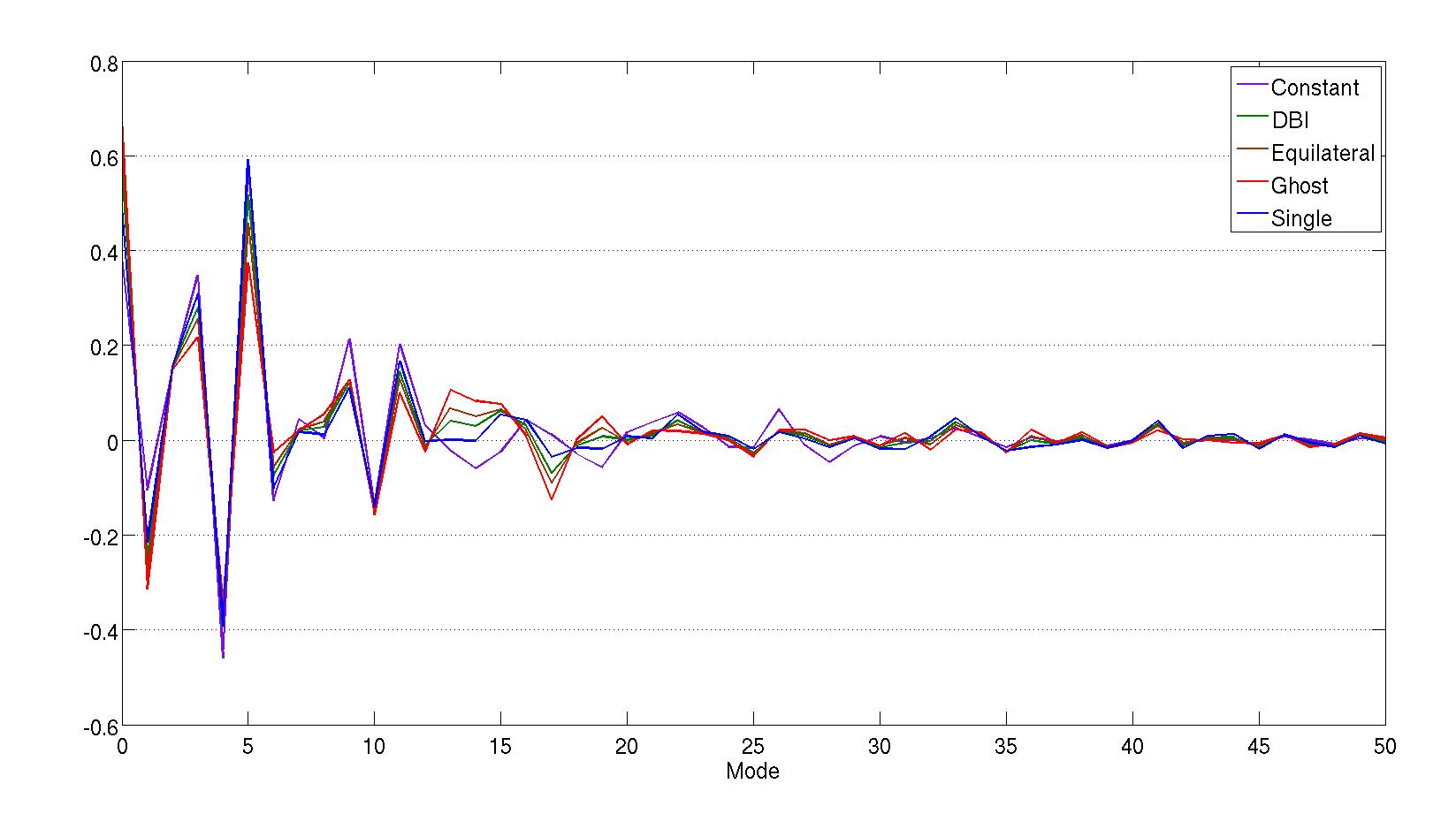}
\includegraphics[width=.9\linewidth, height = 5cm]{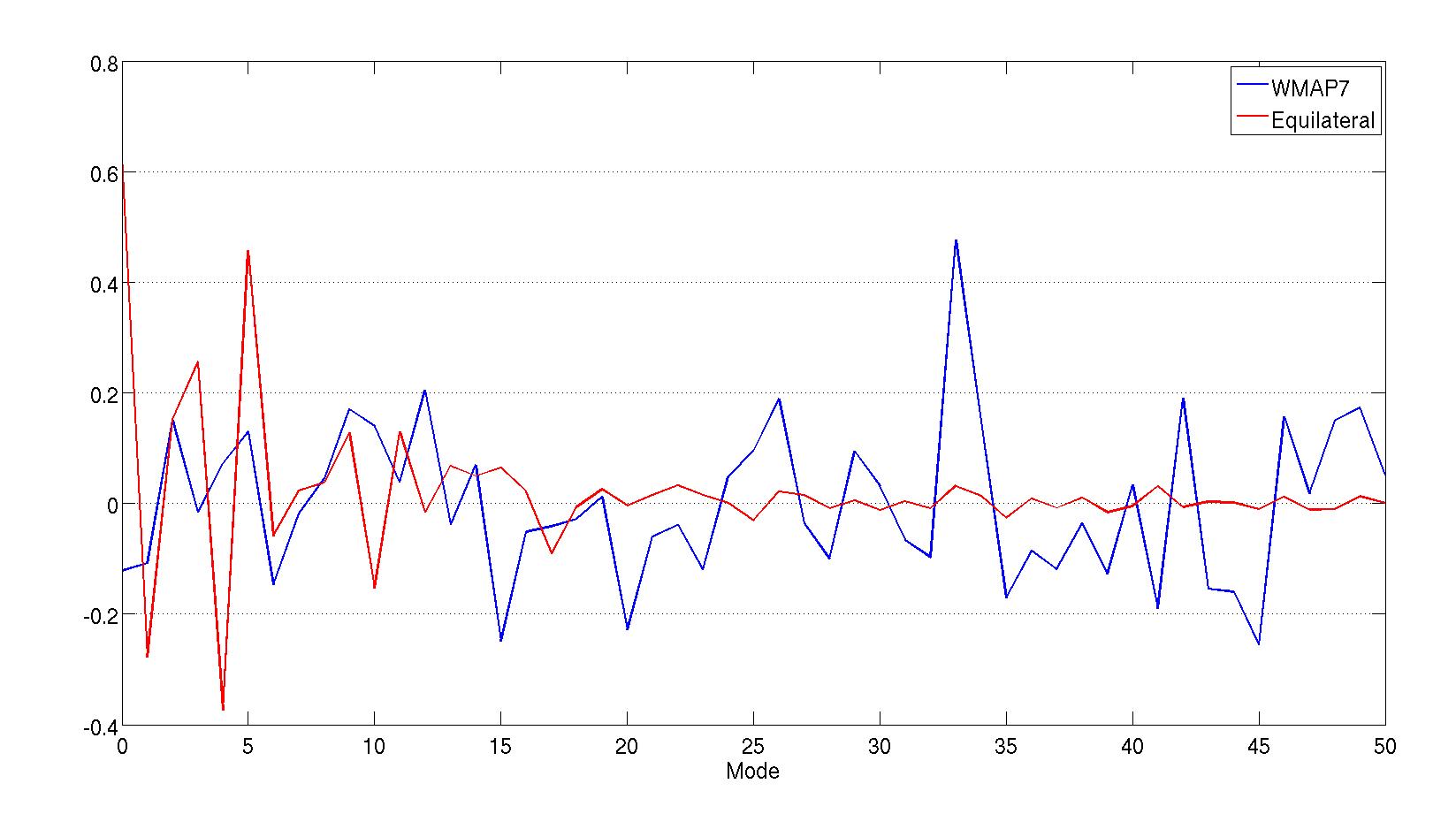}
\caption[]{\small   Equilateral model expansion coefficients $\baRn$ compared between models (top panel) and 
compared with WMAP7 results (lower panel).}
\label{fig:equimodels}
\end{figure}

Bispectra dominated by contributions from nearly equilateral triangle configurations, $k_1\approx k_2\approx k_3$ 
are produced through the amplification of nonlinear effects around the time modes exit the horizon, which can be 
achieved by modifying kinetic terms, as in the DBI model \citep{AlishahihaSilversteinTong2004}, or by explicitly adding higher derivative terms, such as in K-inflation \citep[see, for example,][]{ChenetAl2007}. For DBI inflation, this leads to non-Gaussianity being produced with a shape function of the form \citep{Creminelli2003, AlishahihaSilversteinTong2004}
\eq\label{eq:dbiS}
S(\kall) = \frac{1}{k_1 k_2 k_3 (k_1+k_2+k_3)^2} \[\sum_i k_i^5 + \sum_{i \neq j}\(2 k_i^4 k_j - 3 k_i^3 k_j^2\) 
+ \sum_{i \neq j \neq l}\(k_i^3 k_j k_l - 4 k_i^2 k_j^2 k_l\)\].
\qe
This shape is illustrated in fig.~\ref{fig:equipics}, together with ghost inflation 
\cite{ArkaniHamedetAl2004} and a third distinct single field equilateral shape found in a general 
analysis of such models \cite{ChenetAl2007}.   Note that the generic equilateral shapes are not separable, 
but have been approximated to date using a separable ansatz commonly called the `equilateral model' \cite{CreminellietAl2006}:
\begin{align} \label{eq:equi}
S^{equi}(k_1,k_2,k_3) = \frac{1}{N} \frac{(k_2+k_3 - k_1)(k_3+k_1-k_2)(k_1+k_2-k_3)}{k_1k_2k_3}\,.
\end{align}
Despite the apparent visual differences between these primordial shapes, particularly near the edges of the tetrahedral domain, the resulting CMB bispectra share at least a 95\% or greater correlation  (\citep[see][]{Fergusson:2008ra}).   The CMB  mode decomposition for these models is illustrated in fig.~\ref{fig:equimodels}, showing very similar behaviour to the constant model
also dominated by the acoustic peak coefficients $n=3,4,5$.   The resulting constraints from the modal estimator are:
\eq
\label{eq:fnlequi}
&\hbox{Equilateral:}~~ &\Fnl = 1.90 \pm 23.79 \qquad (\fnl=10.19 \pm 127.38)\,,\\
&\hbox{DBI:}~~~~~~~~~~ &\Fnl = 3.36 \pm 23.86 \qquad (\fnl=17.14 \pm 121.80)\,,\\
&\hbox{Ghost:}~~~~~~~~ &\Fnl = 0.10 \pm 23.68 \qquad (\fnl=0.60~ \pm 139.05)\,,\\
&\hbox{Single:}~~~~~~~ &\Fnl = 5.35 \pm 23.99 \qquad (\fnl=24.56 \pm 110.00)\,.
\qe
Here, the local $\Fnl$ normalisation (\ref{eq:newfnl}) yields much more consistent variances between 
models within the equilateral family than $\fnl$ (as well as values comparable to local and other models).    Note 
that there some variations between the central values of these $\Fnl$ constraints despite the strong correlations 
between these bispectra, because of the different behaviour near the edges where much of the apparent 
WMAP signal is localised.   These  results are consistent with the evolving constraints obtained in the literature 
to date, as shown  in Table~\ref{tab:review}.

Finally, we consider  a separable `orthogonal' shape $S^{\rm orthog}$ which is a constructed from a linear combination of the constant and equilateral shape functions $S^{\rm orthog} \propto S^{\rm equil} -2/3$ (see \cite{MeerburgVanDerSchaarCorasaniti2009, SmithetAl2009}).  The constraint from the mode estimator (\ref{eq:estimatorsum}) then becomes
\eq
\label{eq:fnlortho}
\Fnl^{\rm ortho} = -12.40 \pm 25.02 \,,\qquad (\fnl^{\rm ortho}=-51.42 \pm 103.79)\,,
\qe
which is a less negative result than the latest WMAP7 limit $\fnl^{\rm ortho}=  -199 \pm 104$, but it remains consistent, especially given the lower $\lmax$ employed here. 

\subsection{Flat (trans-Planckian) models}\label{sec:flattened}

\begin{figure}[b]
\centering
\includegraphics[width=0.49\linewidth]{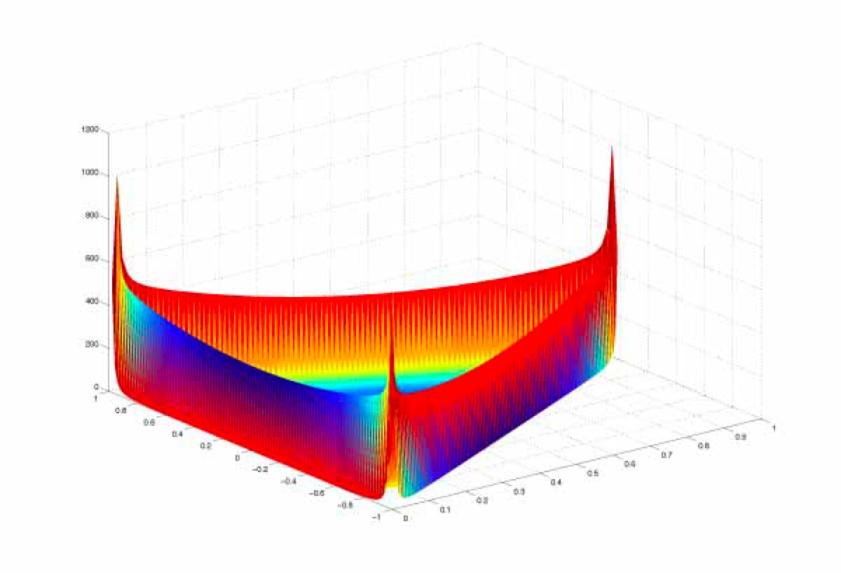} 
\includegraphics[width=.49\linewidth]{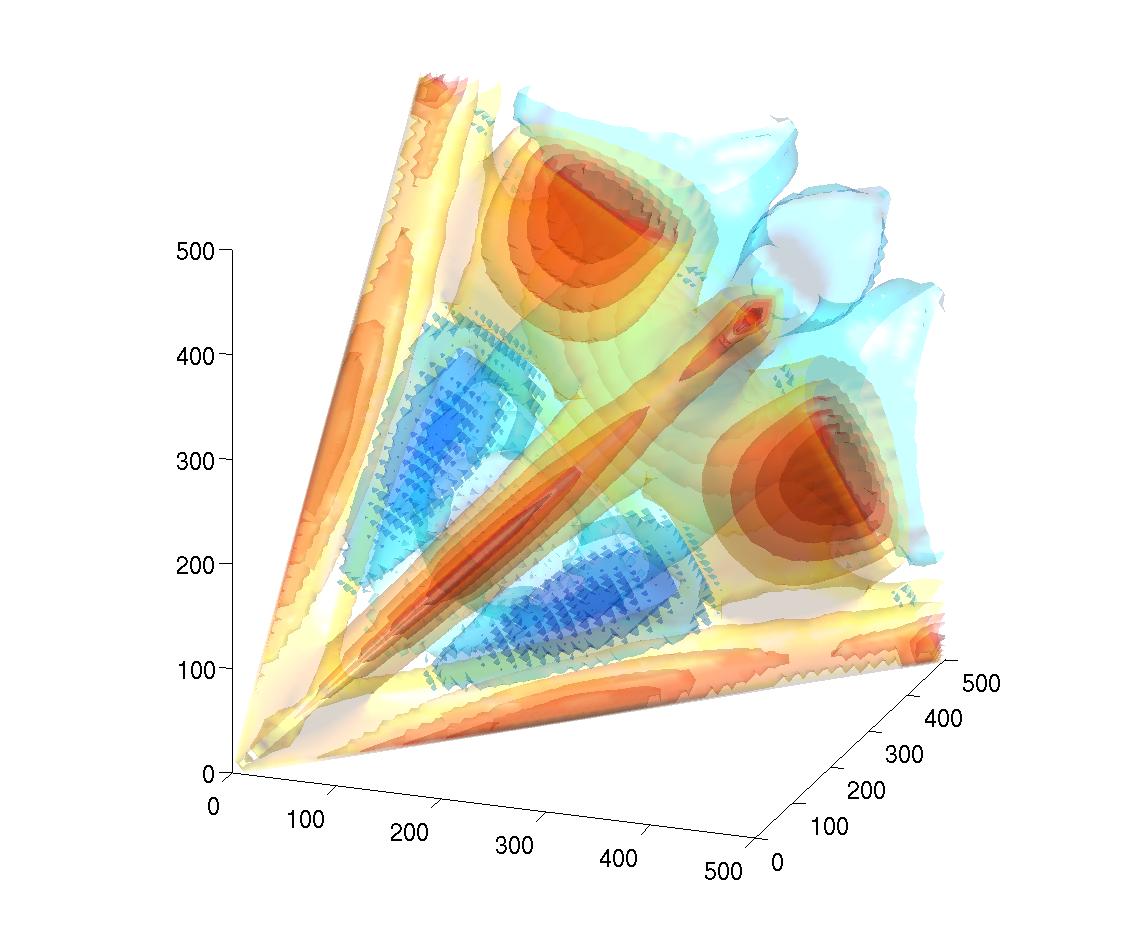}
\caption[Recovered bispectrum]{\small    Flattened model: smoothed primordial shape function (left) and three-dimensional
CMB bispectrum (right) for the flattened model. }
\label{fig:flat}
\end{figure}

\begin{figure}[t]
\centering
\includegraphics[width=.9\linewidth, height = 5cm]{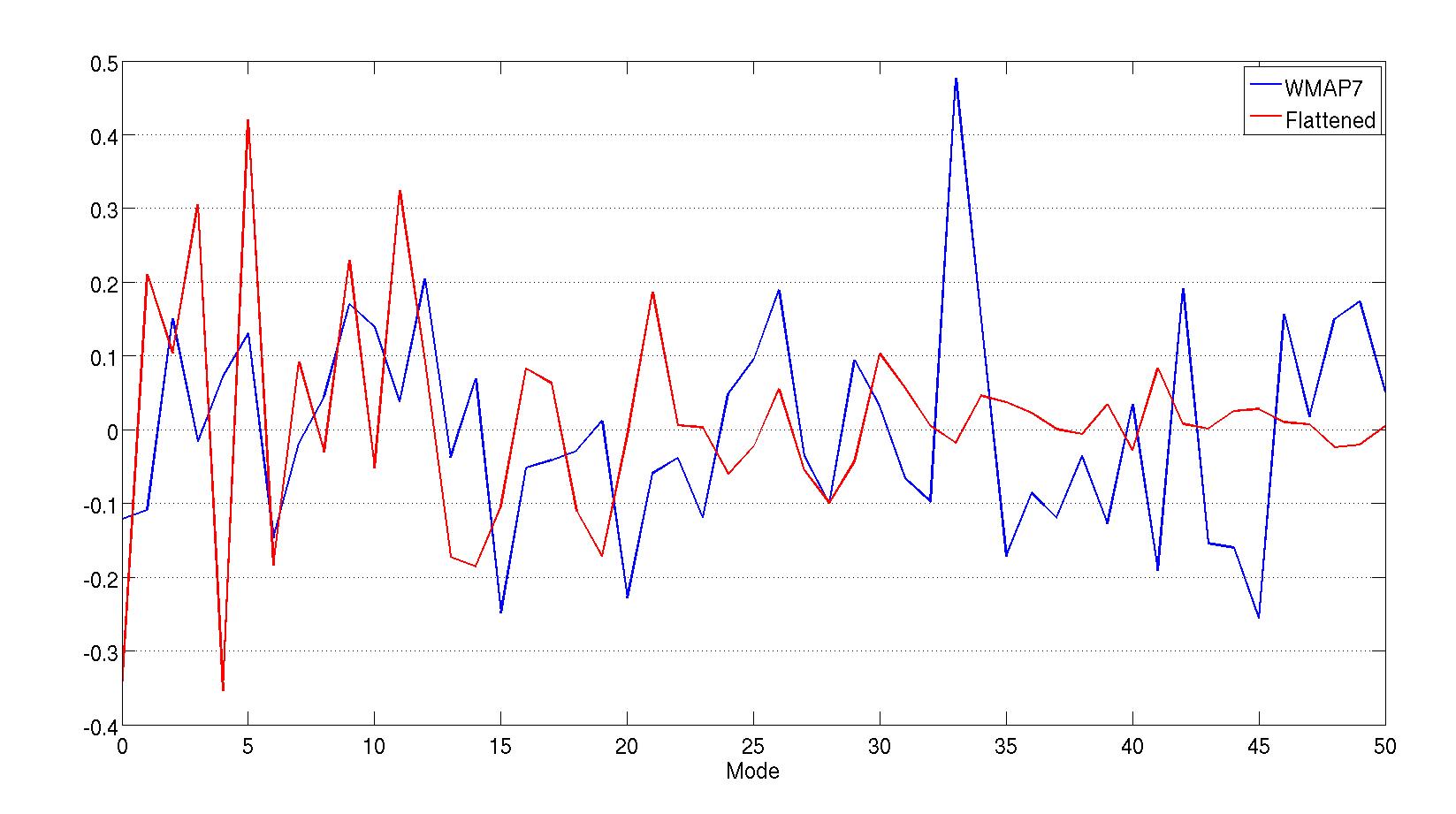}
\caption[]{\small   Flat model mode coefficients compared to WMAP7 mode coefficients. }
\label{fig:flatalpha}
\end{figure}

It is possible to consider inflationary vacuum states which are more general than the Bunch-Davies vacuum, such as an excited Gaussian (and Hadamard) state \citep[][see also discussions in \citealt{ChenetAl2007,MeerburgVanDerSchaarCorasaniti2009}]{HolmanTolley2008}.   Observations of non-Gaussianity in this case might provide insight into trans-Planckian physics. The proposed non-separable shape for the bispectrum is
\eq
\label{eq:flat}
S^{\rm flat}(k_1,k_2,k_3) \propto 6\frac{\(k_1^2+k_2^2 -k_3^2\)}{k_2k_3} +\mbox{2 perms} 
+ 2\frac{k_1^2+k_2^2+k_3^2}{(k_1+k_2-k_3)^2(k_2+k_3-k_1)^2(k_3+k_1-k_2)^2}\,.
\qe
The bispectrum contribution from early times is dominated by flattened triangles, with e.g.\ $ k_3 \approx k_1+k_2$, and for a small sound speed $c_s\ll 1$ can be large. Unfortunately, as the divergent analytic approximation breaks down at the  boundary of the allowed tetrahedron, some form of cut-off must be imposed, as shown for the smoothed shape in fig.~\ref{fig:flat} where an edge truncation has been imposed together with a mild Gaussian filter.  This leads to a degree of predictive uncertainty, but 
the regularisation scheme ensures the primary signal is well-localised on the tetrahedral faces and is quite distinct 
from other separable shapes investigated to dat, including the much more regular orthogonal and folded shapes (refer to ref.~\cite{Fergusson:2008ra} for the specific details).

The resulting CMB spectrum reflects this behaviour with the dominant signal residing near the tetrahedral faces as shown in fig.~\ref{fig:flat}.
Figure~\ref{fig:flatalpha} shows the flat model mode coefficients, which like the local model are only slowly convergent.  Comparing 
the flat model with the coefficients obtained from WMAP, the mode estimator
yields the new constraint:
\eq
\label{eq:fnlflat}
\Fnl = 7.31 \pm 26.22 \qquad (\fnl=3.04 \pm 10.89)\,.
\qe

\subsection{Warm inflation models}

Finally, we consider warm inflation scenarios, that is, nearly scale-invariant models in which dissipative effects play a dynamical role, because these also may produce significant non-Gaussianity \cite{MossXiong2007} (for a review see \cite{BereraMossRamos2009}). Contributions are again dominated by squeezed configurations but with a different more complex shape possessing a sign flip as the corner is approached.  essentially making the signal orthogonal to the local shape.  It can be shown that this makes the warm  and local shapes essentially orthogonal with only a 33\% correlation (see ref.~\cite{Fergusson:2008ra} where the shape function and 
CMB bispectra are discussed).  As with the flat model, uncertainties remain as to the validity of the approximations made as the 
corners and edges of the tetrapyd are approached.    Comparison of the predicted warm bispectrum coefficients $\bbRn$ with the
WMAP data through the modal estimator (\ref{eq:estimatorsum}) yields the constraint 
\eq
\label{eq:fnlwarm}
\Fnl^{\rm warm} = 2.10 \pm 25.83 \qquad(\fnl^{\rm warm}=4.30 \pm 52.83)\,.
\qe
A previous WMAP3 warm inflation analysis obtained a lower central value $\fnl^{\rm warm}=-169\pm 103$ \cite{MossGraham2007}
which is marginally consistent with (\ref{eq:fnlwarm}) at the 95\% confidence level. Probably the most significant difference is 
that the previous analysis did not include a linear term in the estimator (\ref{eq:approxestimator}) to account for noise and masking effects; these corrections are significant here as for the edge-dominated local model.

\section{Implications for non-scaling feature models}\label{sec:feature}
\label{sec:nonscaling}

\begin{figure}[b]
\centering
\includegraphics[width=.85\linewidth, height = 5cm]{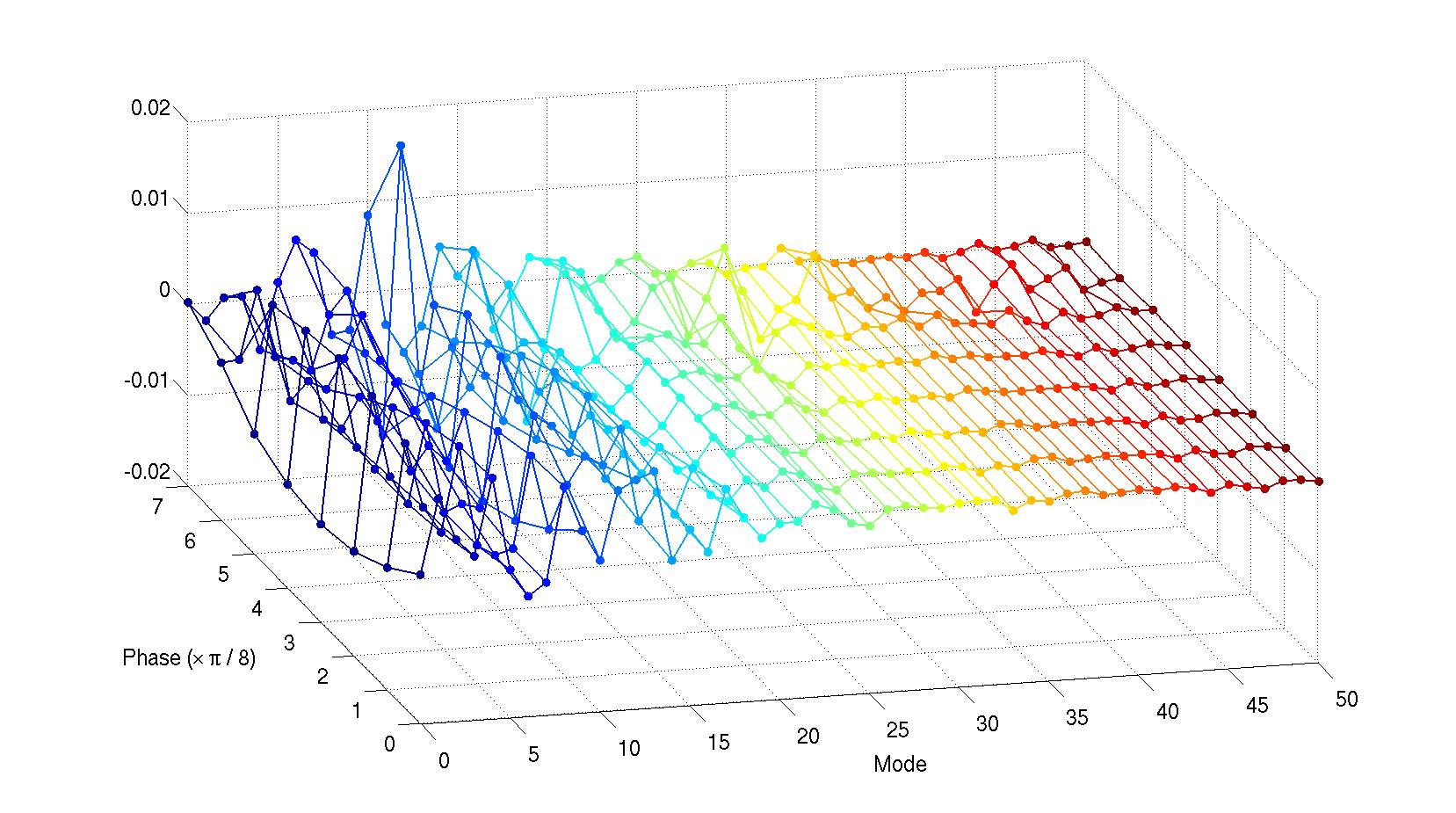}
\includegraphics[width=.85\linewidth, height = 5cm]{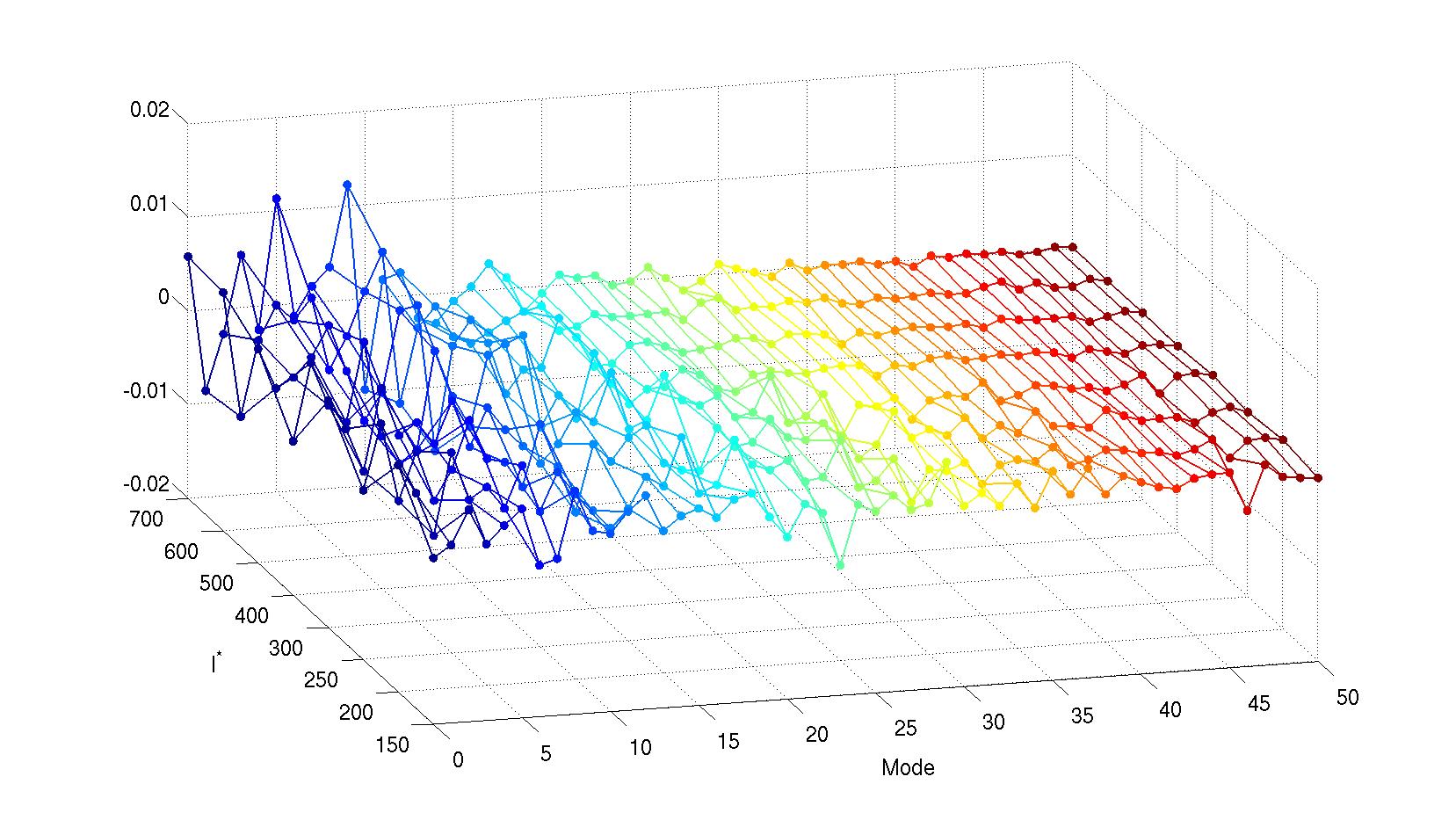}
\caption[]{\small   Feature model coefficients $\baRn$ plotted in two-dimensions by mode number $n$ and as function of 
phase $\phi$ with $l^*=400$ (top panel) and as a function of scale $l^*$ with $\phi=0$ (lower panel).   Note how the 
characteristic $n=3,4,5$ primordial acoustic peak signature is affected (compare with fig.~\ref{fig:constalpha}.}
\label{fig:featurephase}
\label{fig:featurescale}
\end{figure}

\begin{figure}[th]
\centering
\includegraphics[width=.75\linewidth]{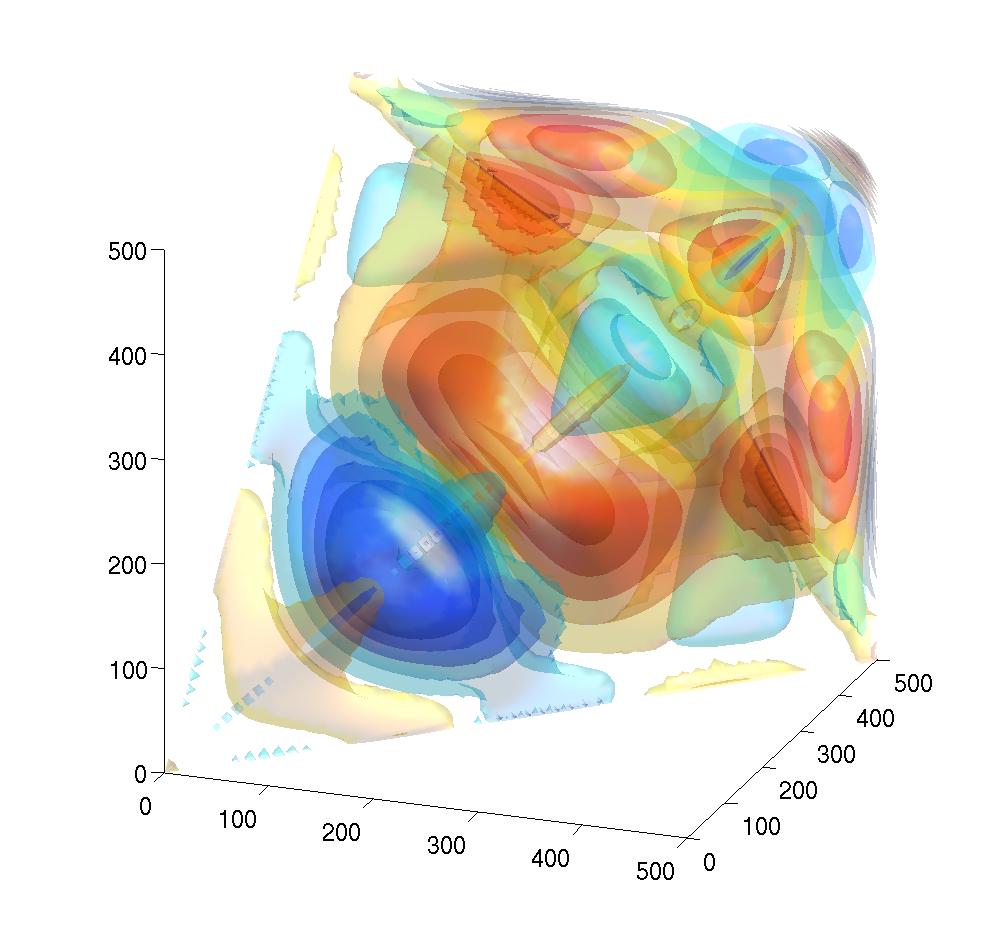}
\caption[]{\small   Three-dimensional CMB bispectrum calculated for the 
feature model   ($l^*=150$, $\phi=0$).  
Note how the scale-dependence of the 
central peaks mimics at some level that observed in the WMAP data. }
\label{fig:featurefit}
\end{figure}

It is possible to produce non-Gaussian signals which are not scale-invariant, such as models with 
a distinct feature in the inflaton potential.   These usually take the form of either a step in the potential (models which have 
a long history, see e.g.\ ref.~ \citep{ChenEastherLim2008}) or those with a small oscillation superimposed onto 
the potential (which have become more popular recently, see e.g.\ ref.~\citep{BeanetAl2008B}.   Two analytic forms for the 
resulting three point functions have been presented in ref.~\citet{ChenEastherLim2008} with the expression we will 
analyse here taking the form 
\begin{align}
\label{eq:feature}
S^{feat}(k_1,k_2,k_3) = \frac{1}{N} \sin\(2\pi\frac{k_1+k_2+k_3}{3k^*} + \Phi\)\,, 
\end{align}
where $k^*$ is the associated with the physical scale of the feature in question and $\Phi$ is an arbitrary phase factor. 
The alternative form with a logarithmic momentum dependence in the $\sin$ argument can be shown to be closely 
correlated with the simpler form (\ref{eq:feature}), certainly on the present domain of study $\lmax =500$.   Previously, 
we studied the shape and CMB bispectrum for a particular feature model (with $k^* \approx l^*/\tau_0$ and $l^*\px400$),  
showing that its non-scaling behaviour made it essentially independent of all the other shapes \cite{Fergusson:2008ra}.  
Such models can have starkly contrasting CMB bispectra as illustrated in fig.~\ref{fig:featurefit}, disrupting
the usual pattern of acoustic peaks which switch from correlation to anticorrelation on multipole scales $l^*$.
Clearly, scale dependent feature models form a distinct category of bispectra beyond the equilateral, local, warm and flat 
families, so searches within WMAP and future data sets are well-motivated.     The 3D bispectrum for a particular feature model is shown in fig.~\ref{fig:featurefit}
demonstrating how such models affect the scale-dependence of the bispectrum  (see fig.~\ref{fig:3dreconWMAP5}).

For the present WMAP7 analysis, we have studied the primordial feature shape (\ref{eq:feature}) over a wide range of 
for which the CMB bispectra that we obtained could be accurately described by our $n=51$ eigenmodes, that is, 
for which we could obtain $>95\%$ convergence to $\blll^{\rm feat}$ for the partial sum (\ref{eq:cmbestmodes}).  This 
restricted the scale parameters in (\ref{eq:feature}) to the range  $l^*\ge 150$, so we studied values $l^* = 150, \,200,\,
250,\,300,\,400,\,500,\,600,\,700$.   For larger values $l^*>700$ the models became highly correlated with the constant
model given that $\lmax=500$.  No such restriction applied to the phase which was studied for each $l^*$ over the full 
domain $0\le \Phi <2\pi$ in $\pi/8$ steps (noting that models separated by $\pi$ are merely anticorrelated).  This 
entailed considerable computational effort calculating 64 distinct CMB bispectra at high accuracy using the robust methods 
previously described elsewhere \cite{FergussonShellard2007}.   The mode coefficients for the $l^*=400$ model are illustrated for the 
different phases in fig.~\ref{fig:featurephase}, demonstrating how the characteristic acoustic peak signal in $n=3,4,5$ can
be modified (compare the constant model fig.~\ref{fig:constalpha}).   The strong dependence of the mode coefficients on the
 different multipole scales $l^*$ (at fixed phase $\Phi = 0$) are shown in fig.~\ref{fig:featurescale}.

\begin{table}[t]
\begin{tabular}{| c | c | c | c | c | c | c | c | c |}
\hline
\backslashbox{\small{\bf Phase}}{\small{\bf Scale}} & 150 &  200 & 250 & 300 & 400 & 500 & 600 & 700\\
\hline
$0$ & $26.93$ & $-37.88$ & $-38.44$ & $-37.88$ & $8.46$ & $13.61$ & $21.01$ & $14.63$ \\
$\pi/8$ & $33.89$ & $-24.90$ & $0.43$ & $-16.65$ & $2.37$ & $9.29$ & $33.78$ & $13.93$ \\
$\pi/4$ & $35.94$ & $-9.16$ & $-32.83$ & $-5.48$ & $-4.00$ & $3.44$ & $-3.17$ & $11.09$ \\
$3\pi/8$ & $31.01$ & $-5.63$ & $-29.79$ & $-27.99$ & $-10.00$ & $-1.10$ & $-5.52$ & $4.75$ \\
$\pi/2$ & $12.45$ & $29.78$ & $-18.75$ & $-31.69$ & $-16.14$ & $0.02$ & $-9.46$ & $-2.42$ \\
$5\pi/8$ & $4.89$ & $43.79$ & $-3.41$ & $-27.43$ & $-24.56$ & $-15.63$ & $-12.37$ & $-7.20$ \\
$3\pi/4$ & $-4.03$ & $48.68$ & $13.17$ & $-17.71$ & $-3.36$ & $-15.63$ & $-14.74$ & $-8.28$ \\
$7\pi/8$ & $-16.37$ & $46.20$ & $28.31$ & $-5.81$ & $-11.52$ & $-15.10$ & $-16.96$ & $-15.45$ \\
\hline
\end{tabular}
\caption{Limits for a selection of feature models in terms of the standard deviation $\Fnl/\Delta\Fnl$ (and for resonant models with which they are highly correlated on these scales).  The typical standard deviation for these results was $\Delta \Fnl = 22.78$ (ranging from 21.90 to 23.43).}
\label{tab:fnllim2}
\end{table}

\begin{figure}[t]
\centering
\includegraphics[width=.9\linewidth]{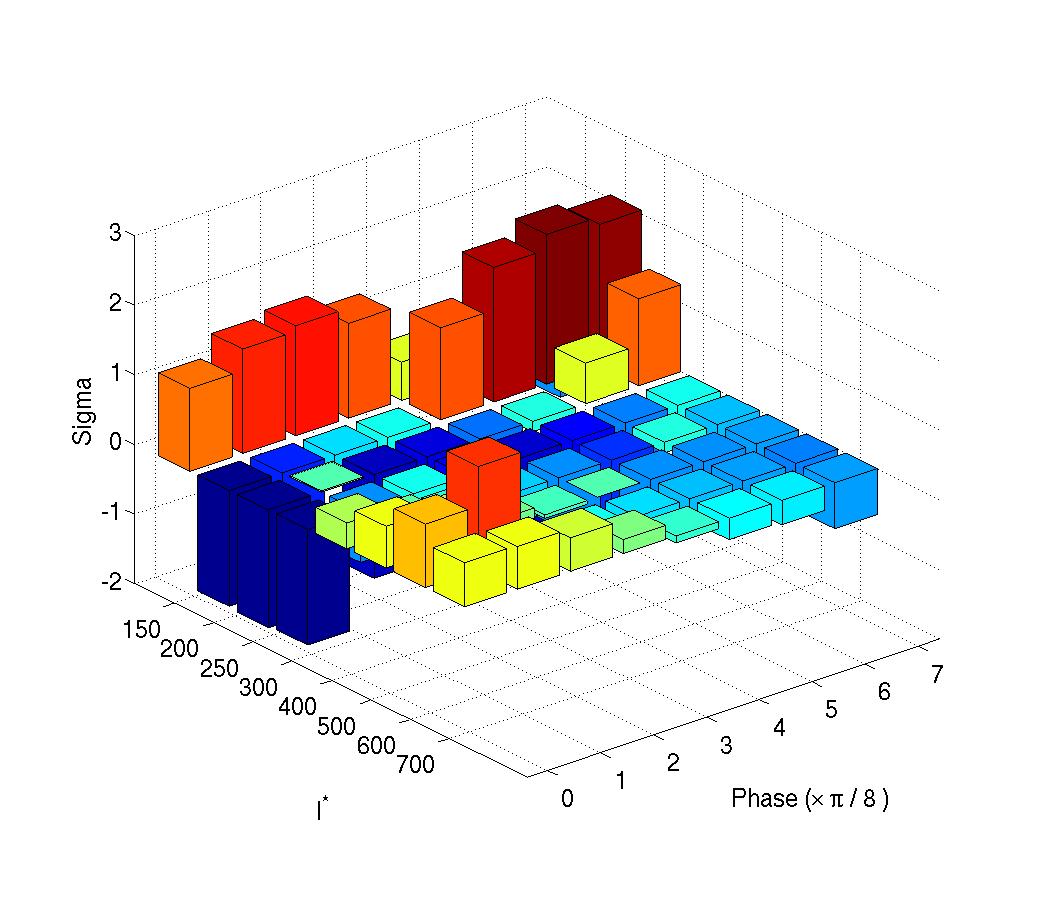}
\caption[]{\small  Significance of feature model bispectra $\Fnl/\Delta\Fnl$ using WMAP data with the modal 
estimator (\ref{eq:estimatorsum}).  This is plotted as a function of the multipole scale $l^*$ and the 
phase of feature models given by (\ref{eq:feature}).}
\label{fig:featuremodels}
\end{figure}

\begin{figure}[t]
\centering
\includegraphics[width=.9\linewidth, height = 5cm]{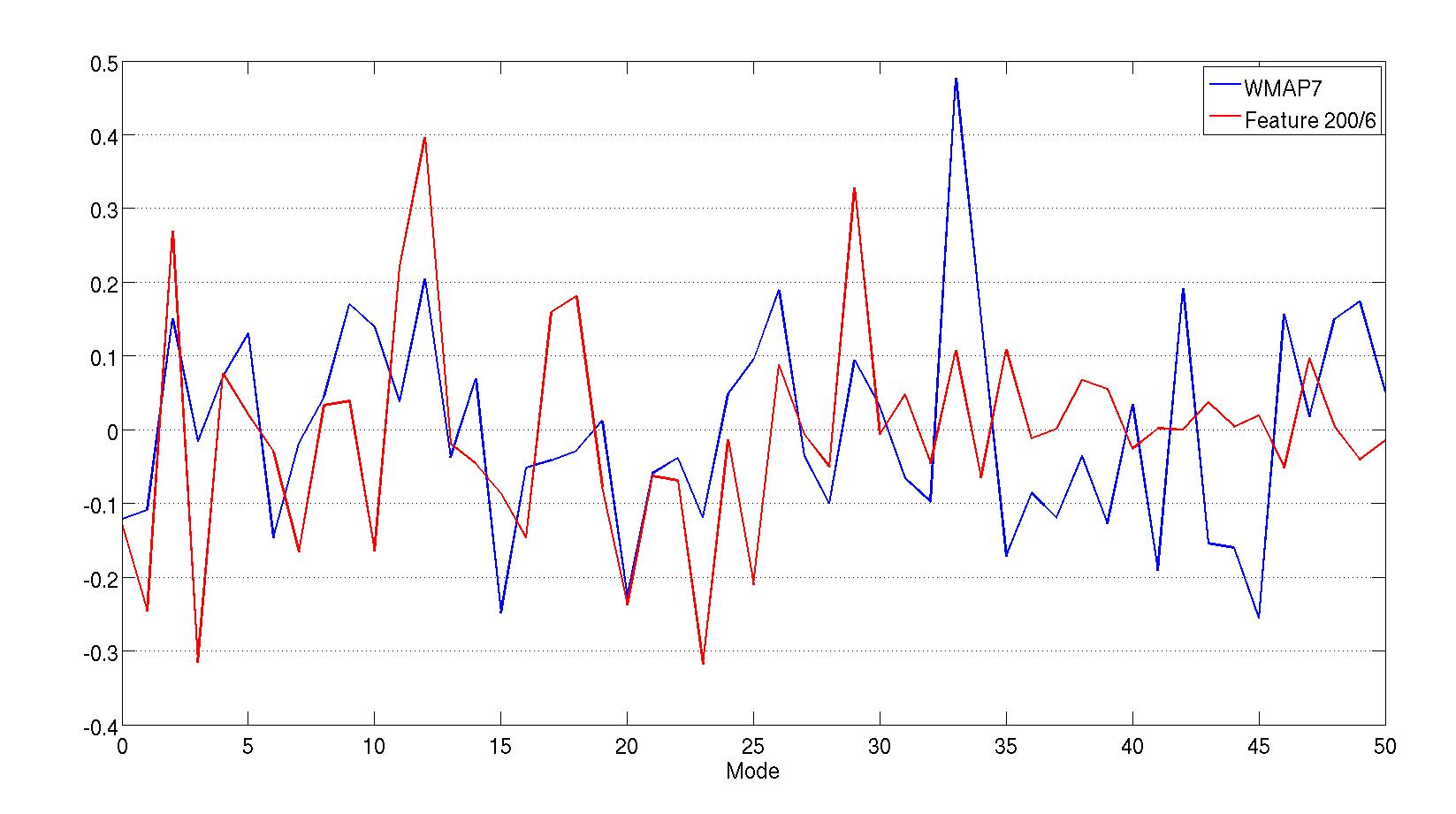}
\caption[]{\small   Best fit feature model coefficients ($l^*=200$, $\phi=3\pi/4$) compared to WMAP7 mode coefficients. }
\label{fig:featalpha}
\end{figure}

Results from the modal estimator for all the feature models investigated are provided in Table~\ref{tab:fnllim2}.
Note that the constraints are given in terms of the normalised quantity $\Fnl$ defined in (\ref{eq:newfnl}), 
since there is no simple generalisation of the primordial normalisation used for $\fnl$ without scale-invariance. 
As before, the variances (given in parentheses) are those obtained for the same set of models 
from 1000 Gaussian simulations.   
The results are illustrated graphically in fig.~\ref{fig:featuremodels} showing the relative significance 
of the central $\Fnl$ values relative to the standard deviation.  
The result with the highest significance is that for the feature model with $l^* = 200$ with a  phase  $\phi = 3\pi/4$
which achieves a $2.15 \sigma$ significance.    There does appear to be more signal as the resolution limit of $l* =150$ is approached.  
 In the overall non-Gaussian analysis, we noted that one anomalous mode n=33 had a 3.39$\sigma$ amplitude.   This mode could be well-correlated with an 
oscillatory model with an effective period below $l^*= 150$ and this is being actively investigated with a higher resolution set of eigenmodes and $\lmax = 1000$ \cite{FergLigShell2012}.  The increased modal signal can probably be associated with the apparent `periodicity' of $l\sim 100-150$ along the diagonal which can be observed in fig.~\ref{fig:3dreconWMAP5}, shorter than the periodicity associated with the transfer functions (e.g.\ see the constant model primary peak in fig.~\ref{fig:const3Dslices}).  Qualitatively the WMAP results look similar to the $(l^*= 150,\; \phi=0)$ feature model plotted in fig.~\ref{fig:featurefit}, though the actual correlation is not that high due to the simplicity of the underlying periodic model (with no overall modulation). 
Nevertheless, the results for the domain of feature 
models investigated $l^*> 150$ remains consistent with the Gaussian hypothesis with no significant detection found on the 
WMAP domain for $l\le 500$.

\section{Towards a measure of the total integrated bispectrum $\barFnl$}\label{sec:totalbisp}
 \label{sec:integratedfnl}
 
 Our focus in this paper has been on recovering the observed bispectrum $\blll$ which contains more 
information than  $\fnlth$ constraints for particular models.  We can also consider blind tests of Gaussianity by considering the quantity
\eq\label{eq:totalbispectrums}
\bar F_{\rm NL}^2 \equiv \sum \bbRn \zeta^{-1}_{np}\bbRp
\qe
which can be interpreted as an ``excess variance" estimator\footnote{As an unambiguous signature of a significant bispectrum we should compare 
$\barFnl$ with the skewness $\gamma_1$ which is 
given by \cite{Regan:2010cn}
\eq\label{eq:skewness}
\gamma_1 \equiv\left \langle \left(\frac{\Delta T}{T}(\hat {\bf n})\right)^3\right\rangle = \frac{1}{4\pi} \sum_{l_i} \hlll ^2\blll\,.
\qe
In principle, the skewness can conspire to vanish even with a non-zero bispectrum $\blll$ because it is not 
positive definite, in contrast to the bispectrum contribution to $\barFnl$.}. If we assume that there is a bispectrum in the data which was not correlated with any of the standard models so $\b_n = f_{\rm NL} \a_n + \b_n^G$ where the superscript $G$ denotes the Gaussian ``noise" then we can calculate
\eq\label{eq:fnl2mean}
\< \bar F_{\rm NL}^2 \> \= f^2_{\rm NL} \(\baRn \zeta^{-1}_{np}\baRp\) + n_{max} \\
&= \(\frac{f_{\rm NL}}{\D f_{\rm NL}}\)^2 + n_{max}
\qe
where we have used $\D f_{\rm NL} = \sqrt{1 / \baRn \zeta^{-1}_{np}\baRp}$. We can also calculate the approximate error:
\eq\label{eq:fnl2var}
\D \bar F_{\rm NL} = \sqrt{ 4 \(\frac{f_{\rm NL}}{\D f_{\rm NL}}\)^2 + 2n_{max} }\,.
\qe
We note, however, 
that (\ref{eq:totalbispectrums}) gives rise to a $\chi^2$-distribution, so we have to take care in assuming Gaussianity for small $\nmax$ and so this expression should been seen as an estimate.
For our present 51 modes the Gaussian variance is $\D \bar F_{\rm NL}\px 11$ and so if there was a specific bispectrum in the data at $\px 4.5\sigma$ then our blind estimator should detect it at $\px 2\sigma$ (though the result would be likely be more insignificantly inconsistent with Gaussianity).

\begin{figure}[t]
\centering
\includegraphics[width=.9\linewidth, height = 7.25cm]{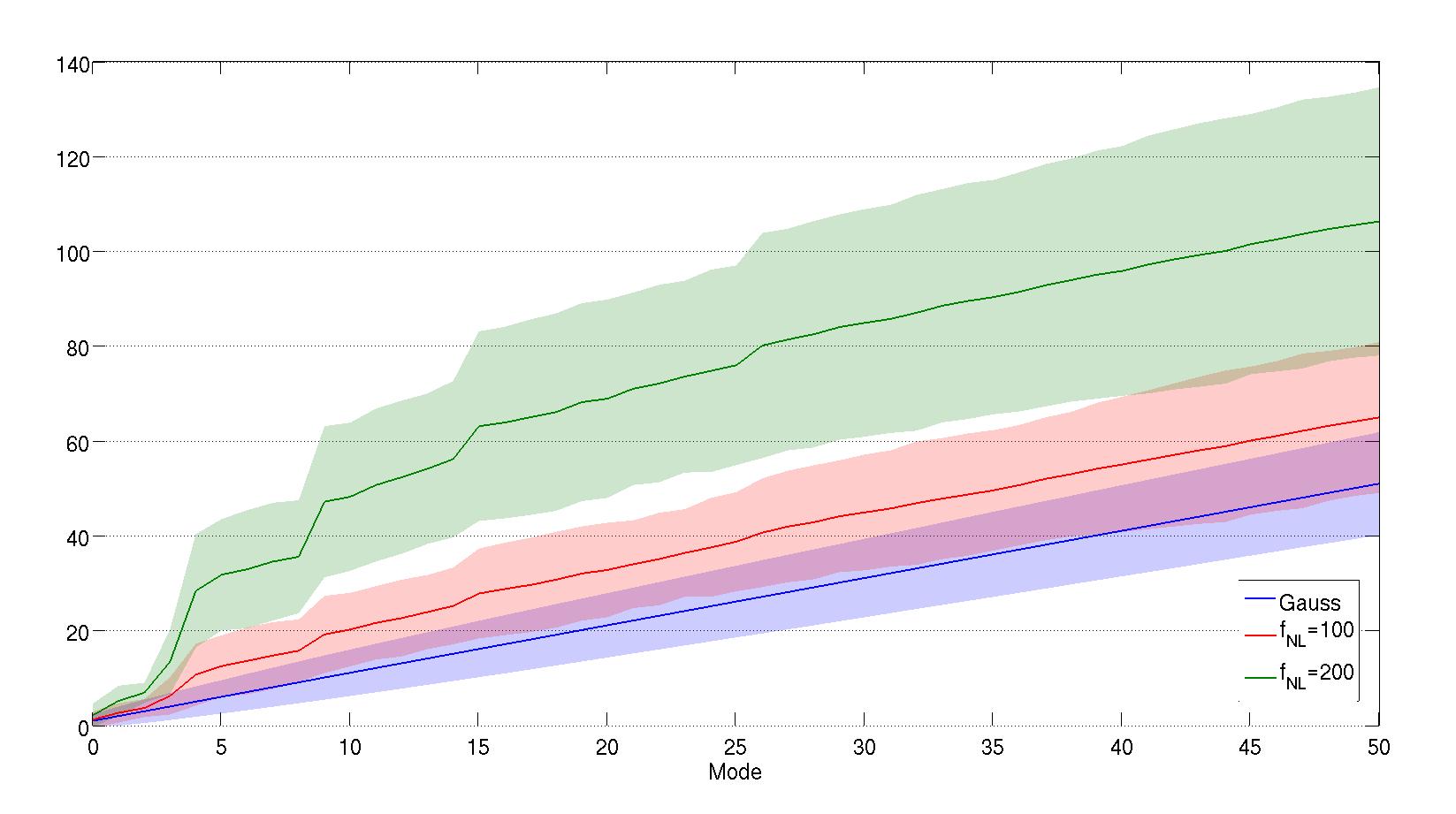}
\caption[]{\small   
Cumulative sum of mode contributions to the total $\barFnl^2$  (\ref{eq:totalbispectrums}) for the local $\Fnl=100$ (red) and $\Fnl=200$ (green)
map simulations compared with Gaussian maps (blue).  The $1\sigma$ variance is shaded around the mean value
obtained from 100 simulations (1000 simulations for the Gaussian case).}
\label{fig:localtotalFnl}
\end{figure}

We estimated $\<{\barFnl^G}{}^2\>$ from 1000 Gaussian simulations and from 100 simulations with various input $\Fnl$. The results are presented in table \ref{tab:FnlRec} and are plotted as cumulative sum of $F_{\rm NL}^2$ 
mode coefficients in  figure \ref{fig:localtotalFnl}. We see the results are close to expected for the Gaussian case but we tend to underestimate the variance for large non-Gaussianity.   
Again, we will further explore the utility of such general modal statistics elsewhere \cite{FergLigShell2012}.

\begin{table}[h]
\centering
\label{tab:FnlRec}
\begin{tabular}{| c | c | c | }
\hline
Input $f_{NL}$ & Mean & StDev\\
\hline
0 & 49.4 (51.0) & 10.7 (10.1) \\
100 & 65.9 (64.1) & 15.2 (12.4) \\
200 & 105.9 (103.4)& 28.4 (17.6) \\
\hline
\end{tabular}
\caption{$\Fnl^{rec}$ as recovered from 100 simulated local maps. Values in brackets are those calculated via the expressions \eqref{eq:fnl2mean} and  \eqref{eq:fnl2var}.}
\end{table}

\begin{figure}[t]
\centering
\includegraphics[width=.9\linewidth, height = 7.25cm]{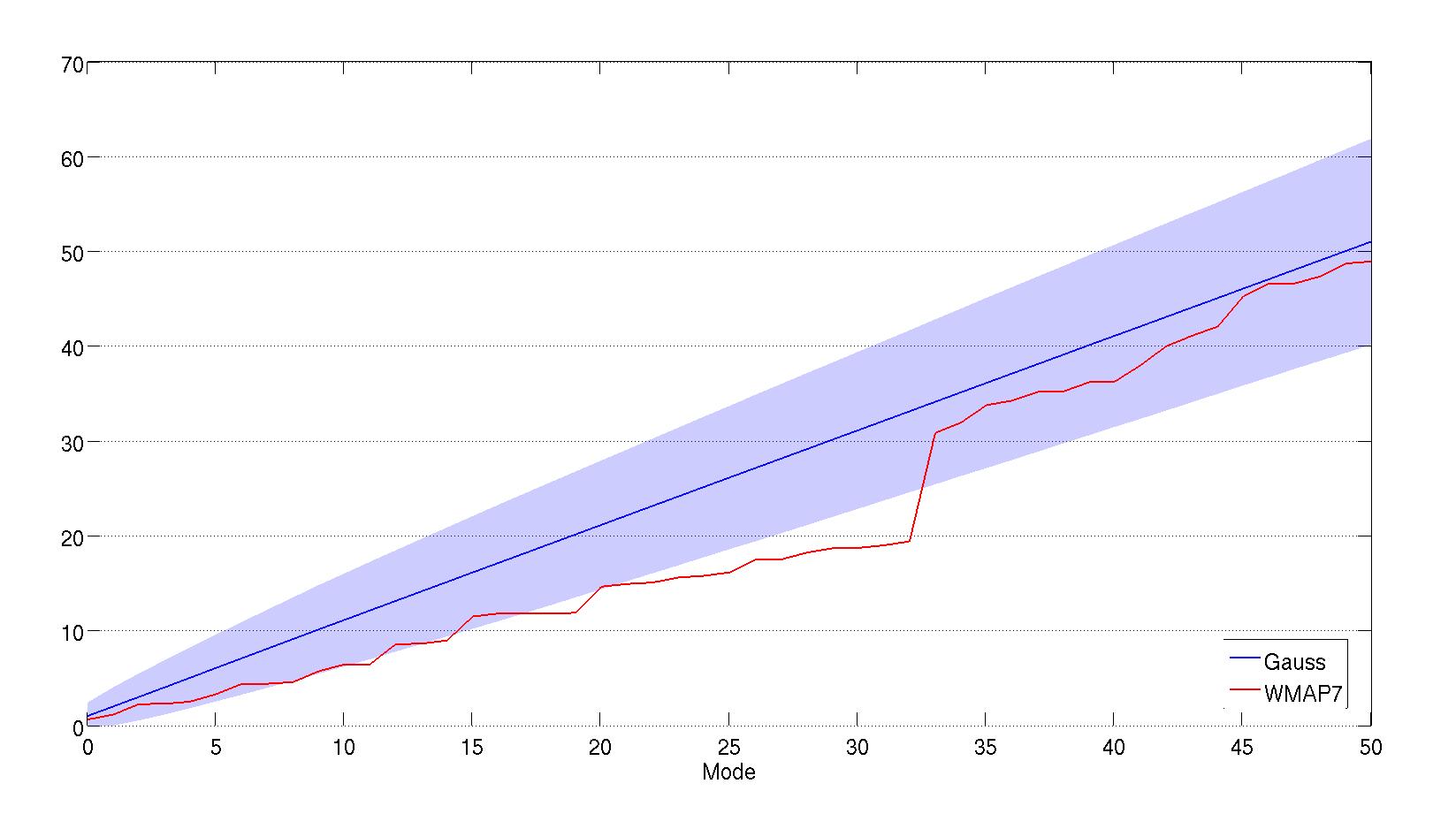}
\caption[]{\small   Cumulative sum of mode contributions to the total bispectrum (\ref{eq:totalbispectrums}) $\barFnl$   for the 
WMAP data compared with Gaussian map simulations as in fig.~\ref{fig:localtotalFnl}.
}
\label{fig:WMAPtotalFnl}
\end{figure}

With the efficacy of the $\barFnl$ statistic established we have also applied it to the WMAP7 data. This yields the unexpected result that  $\bar \Fnl$ obtained from the WMAP7 data is less than that which we we would expect from a typical Gaussian map by slightly over 1$\sigma$ until mode 33 (see the cumulative sum in fig. \ref{fig:WMAPtotalFnl}).  This is somewhat surprising because one would expect a late-time
estimator to be susceptible to foregrounds or other contamination, and it may indicate that the simple WMAP noise model is not adequate.  Nevertheless,  the deviation remains statistically insignificant and the very large mode 33 discussed earlier brings the statistic up to the expected Gaussian mean value.   
However, neglecting this small oddity, the result shown in fig.~\ref{fig:WMAPtotalFnl}  indicates that there is no significant contribution 
to the bispectrum from the first $51$ eigenmodes.   This constrains virtually all smooth scale invariant shapes, as well as 
the feature models we have surveyed.   The only remaining possibility for a bispectrum detection (at the present precision) would then be for oscillatory models with sufficiently high frequencies or bispectra with particularly sharp, or localised, features (i.e. those which 
require $n>51$ for an accurate description).   We have good evidence, therefore, for the null hypothesis that we live in 
a Gaussian universe.

\section{Discussion and conclusions}\label{sec:conclusions}

We have implemented and validated separable mode expansions with a general late-time CMB bispectrum estimator, using it to
investigate a wide range of primordial models with WMAP 7-year data. Notable new constraints include those on non-scaling
feature models, trans-Planckian (flat) models and warm inflation.    The results for nearly scale-invariant models are summarised
in Table \ref{tab:fnllim1}, demonstrating consistency with previous constraints on equilateral and local models. Note that
we adopt a nonlinearity parameter $\Fnl$ normalised to facilitate direct comparison between the local $\fnl$ and any other model.
We found no evidence for significant deviations from Gaussianity for any specific model (at 95\% confidence).
Feature models were surveyed over a wide range of parameters with periodicities above $l^*=150$ and over the
full domain of phase values.   Again, no significant bispectrum detection was made, though given the nature of this
survey some models provide a better a posteriori fit to the data than others.    We note one anomalous 3.4$\sigma$ mode $n=33$ in the bispectrum analysis 
which could correlate with feature or resonant models with $l^*<150$.   The presence of an anomalous signal on related lengthscales could be reproduced with polynomial basis functions with a different ordering.   Together with a higher resolution analysis of the WMAP7 data ($\lmax =1000$) with alternative trignometiric basis functions,  we are investigating the robustness of this signature \cite{FergLigShell2012}.   

More information can be extracted from the mode decomposition of the data than
a few $\Fnl$'s for specific models.   Given that we have constructed a complete orthonormal basis $\barRn$ we can use the
mode coefficients $\bbRn$ to directly reconstruct the full CMB bispectrum using the partial sum (\ref{eq:cmborthmodes}).
We plotted the result for WMAP7 in fig.~\ref{fig:3dreconWMAP5} which, despite its low significance, revealed interesting qualitative features similar to the local model (\ref{fig:localalpha}), but without the periodicity expected from acoustic peaks.   We discussed
a positive-definite measure for  the total integrated bispectrum constructed from the mode coefficients $\barFnl^2 = \sum_n \bbRn{}^2$, which was used to recover $\fnl$ from map simulations in a model independent manner (though
with larger variance).   For WMAP7 data the integrated $\barFnl$ was found to be small and again consistent with
a Gaussian hypothesis.

\begin{table}[t]
\centering
\begin{tabular}{| l | c | c |}
\hline
{\bf Model} & $\Fnl $ & ($\fnl$)\\
\hline
{\bf Constant} & $7.82 \pm 24.57 \; $ & $(30.53 \pm 95.92)$\\
{\bf DBI} & $3.36 \pm 23.86 \; $ & $(17.14 \pm 121.80)$\\
{\bf Equilateral} & $1.90 \pm 23.79 \; $ & $(10.19 \pm 127.38)$\\
{\bf Flat (Smoothed)} & $7.31 \pm 26.22 \; $ & $(3.04 \pm 10.89)$\\
{\bf Ghost} & $0.10 \pm 23.68 \; $ & $(0.60 \pm 139.05)$\\
{\bf Local} & $20.31 \pm 27.64 \; $ & $(20.31 \pm 27.64)$\\
{\bf Orthogonal} & $-12.40 \pm 25.02 \; $ & $(-51.42 \pm 103.79)$\\
{\bf Single} & $5.35 \pm 23.99 \; $ & $(24.56 \pm 110.00)$\\
{\bf Warm} & $2.10 \pm 25.83 \; $ & $(4.30 \pm 52.83)$\\
\hline
\end{tabular}
\caption{Limits for known scale invariant models}
\label{tab:fnllim1}
\end{table}

Despite the absence of any convincing evidence for a statistically significant CMB bispectrum in the present analysis, many
avenues remain open for further investigation using the present methodology.   The late-time modal estimator (\ref{eq:estimatorsum})
can identify any bispectrum whether generated at early times like inflation or sourced since decoupling by
cosmic strings, gravitational lensing, or second-order gravitational effects.   Unlike the primordial estimator, the general mode expansion can also
be used to characterise noise and foregrounds, which need to be identified and subtracted through the linear term
in the estimator (\ref{eq:approxestimator}).   The efficacy of this removal and other validation checks which may affect
a residual local signal will be published shortly \cite{FergLigShell2012}.
Finally, we note again that these methods can be pressed much further with existing and future data, especially from Planck.
The anticipated Planck variance $\Delta \fnl \approx 5$ will substantially improve sensitivity to specific bispectrum shapes,
leaving significant discovery potential available in the near future.
We note also that these separable mode techniques have been adapted for general CMB trispectrum estimation, in principle,
making tractable the investigation of all planar primordial trispectra \cite{Regan:2010cn}.   Analogous methods can also
be applied to modal bispectrum extraction for large-scale structure and in other contexts.    For the time being, however,
this general bispectrum survey uncovers no significant evidence of non-Gaussianity which would undermine
the standard predictions of the simplest models of inflation.

\section{Acknowledgements}

We are very grateful for many informative discussions with Donough Regan, Xingang Chen, Anthony Challinor 
and Alessandro Renzi.  Simulations were performed on the COSMOS supercomputer (an Altix 4700) which is funded by 
STFC, HEFCE and SGI.   We are particularly grateful for computer support from  Andrey Kaliazin.  
 JRF, ML and EPS were supported by STFC grant ST/F002998/1 and the 
Centre for Theoretical Cosmology.	 EPS is grateful for the hospitality
of the Arnold Sommerfeld Centre and the Universe Excellence
Cluster in Munich.

\bibliography{BispectrumII}

\end{document}